\documentclass[twocolumn,times,numberedappendix]{aastex61}

\usepackage{bm,color,ulem}

\newcommand{\pr}[1]{\ensuremath{\left(#1\right)} }
\newcommand{\pd}[2]{\ensuremath{\frac{\partial #1}{\partial #2} }}
\newcommand{\pf}[2]{\ensuremath{\left( \dfrac{#1}{#2} \right)} }
\newcommand{\ave}[1]{\ensuremath{\left\langle#1\right \rangle} }
\newcommand{\aved}[1]{\ensuremath{\ave{\ave{#1}}}}
\newcommand{\dfrac}[2]{ {\displaystyle\frac{#1}{#2}} }
\newcommand{\eqref}[1]{Equation (\ref{#1})}

\newcommand{\vAz}{\ensuremath{  v_{{\rm A}z } }}
\newcommand{\vAzz}{\ensuremath{  v_{{\rm A}z 0} }}
\newcommand{\vAzero}{\ensuremath{  v_{{\rm A}0} }}

\newcommand{\Alfven}{Alfv\'en }

\defcitealias{Okuzumi2015The-Nonlinear-O}{OI15}
\defcitealias{Mori2016Electron-Heatin}{MO16}

\def\figscale {1.0}

\begin{document}

\title{Electron Heating and Saturation of Self-regulating Magnetorotational Instability in Protoplanetary Disks}
\shorttitle{Electron Heating in MRI}

\shortauthors{Mori et al.}
\def\myemail{mori.s@geo.titech.ac.jp}
\def\titech{Department of Earth and Planetary Sciences, Tokyo Institute of Technology, Meguro-ku, Tokyo, 152-8551, Japan}
\def\riken{RIKEN Advanced Institute for Computational Science, Chuo-ku, Kobe, Hyogo, 650-0047, Japan}
\def\nagoya{Department of Physics, Nagoya University, Chikusa-ku, Nagoya, 464-8601, Japan}

\author{Shoji Mori}                
\affiliation{\titech}
\author{Takayuki Muranushi} 
\altaffiliation{Deceased 2017 July 11} 
\affiliation{\riken}  
\author{Satoshi Okuzumi}      
\affiliation{\titech}
\author{Shu-ichiro Inutsuka}  
\affiliation{\nagoya}

\correspondingauthor{Shoji Mori}
\email{\myemail}

\begin{abstract}
Magnetorotational instability (MRI) has a potential to generate the vigorous turbulence in protoplanetary disks, 
although its turbulence strength and accretion stress remains debatable because of the uncertainty of MRI with low ionization fraction. 
We focus on the heating of electrons by strong electric fields which amplifies nonideal magnetohydrodynamic effects.
The heated electrons frequently collide with and stick to dust grains, which in turn decreases the ionization fraction and is expected to weaken the turbulent motion driven by MRI. 
In order to quantitatively investigate the nonlinear evolution of MRI including the electron heating, we perform magnetohydrodynamical simulation with the unstratified shearing box.
We introduce a simple analytic resistivity model depending on the current density by mimicking resistivity given by the calculation of ionization.
Our simulation confirms that the electron heating suppresses magnetic turbulence when the electron heating occurs with low current density. 
We find a clear correlation between magnetic stress and its current density, which means that the magnetic stress is proportional to the squared current density. 
When the turbulent motion is completely suppressed, laminar accretion flow is caused by ordered magnetic field. 
We give an analytical description of the laminar state by using a solution of linear perturbation equations with resistivity.  
We also propose a formula that successfully predicts the accretion stress in the presence of the electron heating.
\end{abstract}

\section{Introduction}
Magnetorotational instability (MRI) has a potential to generate vigorous turbulence in protoplanetary disks.
The turbulent viscosity made by the MRI can explain the accretion rate suggested by observation \citep[e.g.,][]{Hawley1995Local-Three-dim,Flock2011Turbulence-and-}.
That is why MRI has been expected to be a mechanism generating disk turbulence in most research of the protoplanetary disks.
Previous studies have been investigated how MRI turbulence in the disks significantly affects the planetesimal formation. 
For examples, the vigorous MRI turbulence causes diffusion of the dust condensed region \citep{Carballido2005Diffusion-coeff,Fromang2006Dust-settling-i,Fromang2009Global-MHD-simu,Turner2010Dust-Transport-,Zhu2015Dust-Transport-} and the collisional fragmentation of grains \citep{Carballido2010Relative-veloci}.
The disk turbulence is important for both of the disk evolution and planetesimal formation.

However, MRI growth and generation of vigorous magnetic turbulence need the disk to be sufficiently ionized.
Decoupling between the gas and magnetic fields due to the low ionization fraction causes the nonideal magnetohydrodynamic (MHD) effects, such as ohmic dissipation, Hall effect, and ambipolar diffusion.
The nonideal MHD effects can stabilize MRI \citep[e.g.,][]{Fleming2000The-Effect-of-R,Sano2002bThe-Effect-of-t,Bai2011Effect-of-Ambip,Bai2013bWind-driven-Acc,Kunz2013Magnetic-self-o,Simon2015Magnetically-dr}.
The nonideal MHD effects strongly depend on the ionization fraction.
Therefore, it is essential to understand ionization state in the disk to determine the efficiency of MRI and the strength of the resulting turbulence. 

Although the theoretical estimate of the turbulence strength in a disk still have large uncertainties, 
recent disk observations found indirect evidence of the turbulence strength.
The disk around HL Tau, which is thought to be typical protoplanetary disks surrounding T Tauri stars, is observed by ALMA observatory, and then the significantly detailed figure is unveiled with the high spatial resolution \citep{ALMA-Partnership2015The-2014-ALMA-L}.
The disk has many axisymmetric rings and gaps approximately within 100 AU from the star. 
\citet{Pinte2016Dust-and-Gas-in} reproduced the similar observational image with the radiative transfer simulation and obtained the dust and gas properties.
According to the paper, such a clear gap requires for the dust disk to be geometrically thin, which means the weak turbulence as Shakura-Sunyaev alpha parameter $\alpha \la$ a few $10^{-4}$ \citep{Shakura1973Black-holes-in-}.
Moreover, \citet{Flaherty2015Weak-Turbulence} and \citet{Flaherty2017A-Three-Dimensi} observed a disk around A-type star, HD163296, and obtained the spectral map that limits on non-thermal gas velocity dispersion which is mainly due to turbulent motion.  
\citet{Flaherty2017A-Three-Dimensi} constrained that the velocity dispersion is less than $\sim 0.04$ times the sound speed which corresponds to $\alpha \la 10^{-3}$ around midplane.
The value is one order of magnitude less than typical $\alpha$ value of fully developed MRI turbulence $\alpha \sim 10^{-2}$.
The direct imaging observation of HD163296 by \citet{Isella2016Ringed-Structur} which observed multiple gaps also suggested weak turbulence from gap width and depth relation, assuming presence of planet in the gaps.
These disk observations show a new problem of how such weak turbulence is formed.

In this paper, we investigate the effect of electron heating on the MRI.
The electron heating is one of the consequences of resistive MHD and has a potential to suppress MRI via changing ionization balance.
MRI generates not only magnetic fields but also electric fields in the comoving frame of the gas. 
The electric fields induced by the MRI heat charged particles, in particular electrons, in the gas due to collision with gas particles \citep{Inutsuka2005Self-sustained-}.
The heated electrons are efficiently removed from the gas phase because they frequently collide with and stick to dust grains \citep[][hereafter OI15]{Okuzumi2015The-Nonlinear-O}. 
Therefore, the electron heating causes a decrease in the ionization fraction, which amplifies the nonideal MHD effects suppressing MRI.
Since the electron heating takes place after MRI sufficiently grows, the nonideal MHD effects amplified by the electron heating can change the picture of MRI behavior even in sufficiently ionized region.
Our previous study \citep[][hereafter MO16]{Mori2016Electron-Heatin} investigated the region in protoplanetary disks where the electron heating influences MRI. 
We showed that this suppression mechanism becomes important even in outer regions of protoplanetary disks that retain abundant small dust grains.
Since the MRI growth leads to suppress the MRI by itself in the presence of electron heating, the saturated turbulent motion would be weaker than the one of fully developed MRI turbulence. 
\citetalias{Mori2016Electron-Heatin} also estimated the accretion stress of magnetic turbulence by using a scaling relation between the magnetic stress and the current density, and suppose that the accretion stress suppressed by the electron heating can be reduced by more than an order of magnitude.

How much the electron heating suppress MRI is still unclear, although the possibility of occurrence of electron heating in the disks has been investigated.
The estimation of turbulence strength in \citetalias{Mori2016Electron-Heatin} is based on the scaling relation that has not been verified.
In order to confirm the effectiveness for electron heating to suppress magnetic turbulence, accretion stress in the presence of the electron heating should be investigated quantitatively.

Our goal in this work is to quantify the effect of the electron heating on MRI with a numerical simulation.
We perform MHD simulations where the suppression of the electric resistivity due to electron heating is modeled by a simple analytic function.  
Furthermore, we propose a formula that reproduces the Maxwell stress obtained from the simulation, which can be used to take into account the effect of the electron heating on the disk evolution.
As a first step, we neglect ambipolar diffusion and the Hall effect, focusing on how the Ohmic resistivity increasing with the electric field strength affects the saturated state of MRI.
In addition, although strong electric fields do not only heat electrons but also ions, we also neglect the ion heating which requires much higher electric field strength than electron heating \citepalias{Okuzumi2015The-Nonlinear-O}.

The plan of this paper is as follows.
In Section \ref{sec:num-set}, we present the numerical setup and procedure in our simulations.
In Section \ref{sec:sim-results}, we then show some results and present the interpretations.
In Section \ref{sec:der-pred-form}, we analytically derive a relation between current density and Maxwell stress.
In Section \ref{sec:Summary}, we summarize this paper and discuss implications for dust diffusion in protoplanetary disks.

\section{Method}\label{sec:num-set}
\subsection{Numerical Method}
We perform MHD simulations with a unstratified local shearing box, using Athena, an open source MHD code which uses Godunov's scheme \citep{Stone2008Athena:-A-New-C, Stone2010Implementation-}.
We adopt a local reference frame ($x$, $y$, $z$) corotating with the Keplerian flow at a fiducial distance $r_0$ from the central star.
The coordinates $x$, $y$, and $z$ refer to the radial, azimuthal, and vertical distances from the corotation point, respectively.
Neglecting curvature and vertical gravity, the MHD equations in this local coordinate system can be written as
\begin{eqnarray}
   \pd{\rho}{t} +  \nabla \cdot \left(\rho \bm{v} \right) &=& 0  \label{basicEqDensity} \ , \\
  \pd{\bm{v}} {t} + \pr{\bm{v}  \cdot \nabla} \bm{v} &=& - 2 \bm{\Omega} \times  \bm{v} +3\Omega^2 \bm{x}  \nonumber \\ && - \frac{1}{\rho}\nabla\pr{P + \frac{B^2}{8\pi}}  
    + \frac{1}{4\pi\rho} \pr{  \bm{B} \cdot\nabla\bm{B}  }  \label{basicEqVelocity} \ ,  \\
  \frac {\partial \bm{B}}{\partial t} &=& - c \nabla \times \bm{E} \label{basicEqMagnet} \ , 
\end{eqnarray}
where $ \bm{v} $ is the gas velocity, $\rho$ is the gas density, $P$ is the gas pressure, $\Omega$ is the angular velocity at radius $r_0$, $ \bm{B} $ is the magnetic field, $ \bm{E} $ is the electric field, and $c$ is the speed of light.
In this paper, we assume isothermal fluid and use the isothermal equation of state for an ideal gas,
\begin{eqnarray}
  P = c_{\rm s}^2 \rho \ ,       \label{isothermalEqOS}
\end{eqnarray}
where $c_s$ is the sound speed of isothermal gas and constant.
The electric field ${\bm E}$ in this reference frame is related to the electric field $\bm{E}'$ in the comoving frame of the gas, 
\begin{equation}\label{eq:Lol-E}
	 \bm{E}  = \bm{E}'  - \frac{1}{c}  \bm{v}  \times  \bm{B}  \ ,
\end{equation}
by the Lorentz transformation in the limit of small velocity.
To close the system of equations, we employ the Ohm's law,
\begin{equation}\label{eq:Ohm}
	\bm{J} =\frac{c^{2}}{4 \pi \eta(E')}  \bm{E}'  \  ,
\end{equation}
where $\bm{J}=(c/4\pi) \nabla \times \bm{B}$ is the current density.
In this study, we assume that the electric resistivity $\eta$ depends on the amplitude of the electric field strength, $E' = |{\bm E}'|$, which is the case when electron heating changes the ionization fraction.  

The dependence of $\eta$ on $E'$ was investigated in \citepalias{Okuzumi2015The-Nonlinear-O}.
\citetalias{Okuzumi2015The-Nonlinear-O} calculated the ionization fraction from the ionization equilibrium including two important effects of plasma heating, i.e., the amplification of plasma adsorption onto dust grains and impact ionization by energetic plasma.
The amplification of plasma adsorption decreases plasma abundance, while the impact ionization increases plasma abundance.
They showed that the amplification of plasma adsorption occurs at lower $E'$ than impact ionization.
In this work, we focus only on the amplification of plasma adsorption amplified by the electron heating and neglect impact ionization.

\begin{figure}[t]
\centering
 \includegraphics[width=1.0\hsize,clip]{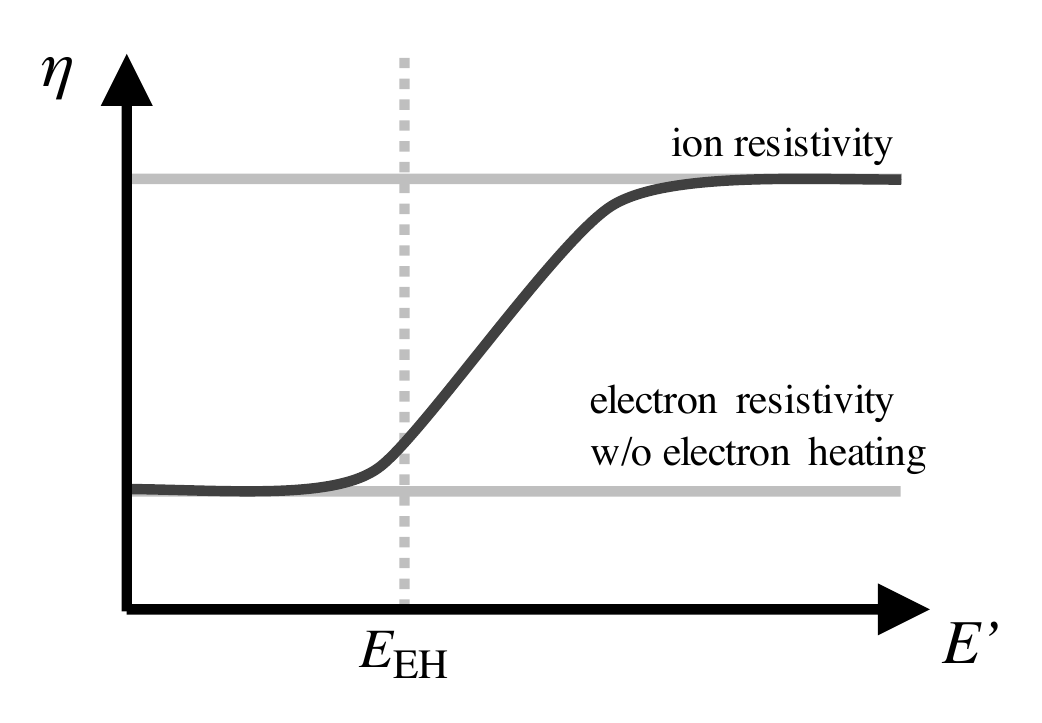}
\caption{ 
Schematic diagram of dependence of resistivity as a function of $E'$ in \citetalias{Okuzumi2015The-Nonlinear-O} that includes amplification of the dust adsorption by the electron heating.  
The dominant charge careers change from electrons to ions with increasing $E'$ due to reduction of electron abundance by the electron heating. 
}
 \label{fig:J-Emodel}
\end{figure}

In this paper, we use an analytical resistivity model that mimics the behavior of $\eta$ as a function of $E'$ due to the electron adsorption which is based on \citetalias{Okuzumi2015The-Nonlinear-O}. 
Figure \ref{fig:J-Emodel} shows a schematic picture of our resistivity model.
Effective electric resistivity is determined by the smaller of the electron and ion resistivity.
The horizontal gray lines show electron and ion resistivity in the case without electron heating.
The critical electric field strength $E_{\rm EH}$ shows the threshold of electric field strength where electron heating occurs.
For $E' > E_{\rm EH}$, the resistivity increases with increase of $E'$ due to dust adsorption of heated electrons.
When $E'$ is so small that electron heating does not work, i.e. $E' \ll E_{\rm EH}$, electron resistivity is much smaller than ion resistivity.
In the case, the effective resistivity is equal to electron resistivity without electron heating which is constant.
On the other hand, the electron resistivity in $E' > E_{\rm EH}$ increases with increases of $E'$ because electron abundance decreases due to electron heating.
In this case, the effective resistivity also increases.
At $E' \gg E_{\rm EH}$, electron resistivity is larger than ion resistivity, and therefore the effective resistivity is determined by ion resistivity and constant.

In this work, we focus only on the resistivity increasing by the electron heating but does not address an instability of electric fields caused by negative differential resistance, $dJ/dE < 0$ \citepalias[see Section 6.1 in][]{Okuzumi2015The-Nonlinear-O}.
In this work, the gradient of $\eta$ to $E'$ is modified to be shallower than the resistivity given in \citetalias{Okuzumi2015The-Nonlinear-O}.
In order to satisfy $dJ/dE =1/(d(\eta J)/dJ) > 0$, the power-law index of $\eta$ to $J$ is taken to be larger than $-1$.

\begin{figure}[t]
\centering
 \includegraphics[width=0.9\hsize,clip]{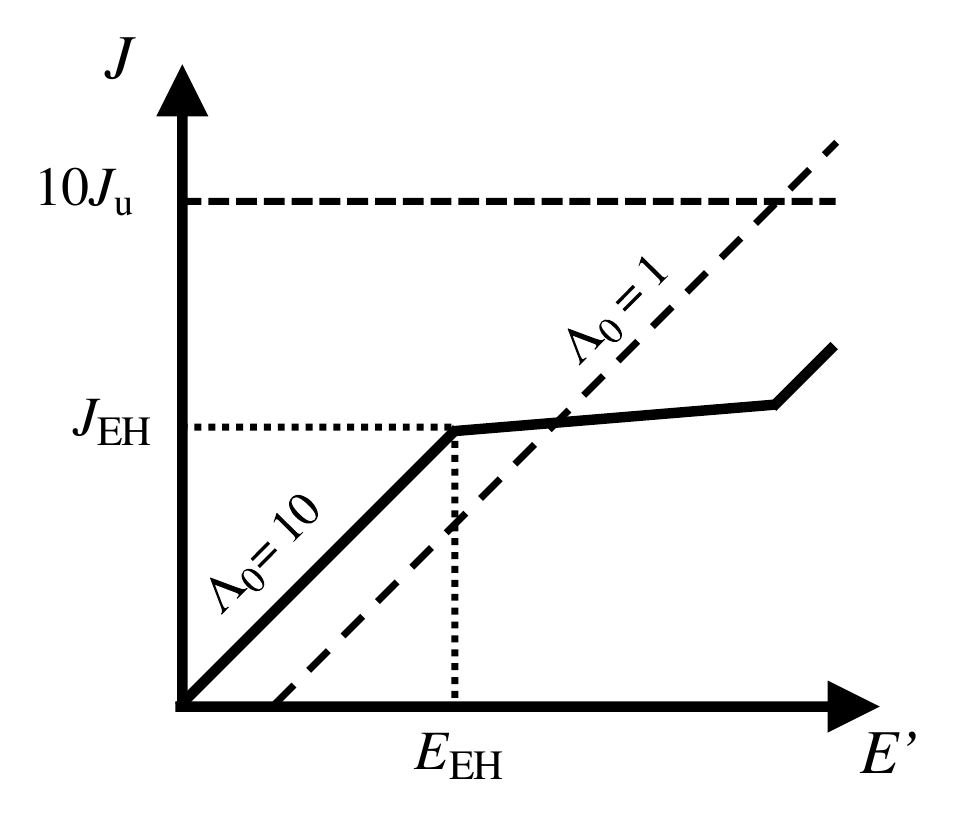}
\caption{
Schematic diagram of the simplified $J$--$E'$ relation that we use in this paper.
We take the initial Elsasser number to be $\Lambda_{0}= 10$ in the fiducial model.
$J_{\rm EH}$ is the current density at which electron heating sets in, and horizontal dashed line of $10 J_{\rm u}$ approximately corresponds to the current density of fully developed MRI turbulence \citep{Muranushi2012Interdependence}, which means the maximum current density.
The simulations are performed with varying different values of $J_{\rm EH}$ below $10 J_{\rm u}$.
}
 \label{fig:J-Esketch}
\end{figure}

Imitating the $J$--$E'$ relation of \citetalias{Okuzumi2015The-Nonlinear-O} including electron heating,
we give the simple analytical resistivity model where the resistivity increases with an increase of $E'$ or $J$.
In Figure \ref{fig:J-Esketch}, we show a schematic diagram of $J$--$E'$ relation including our resistivity model.
The resistivity $\eta$ is written as 
\begin{equation}\label{eq:set-def-eta}
\eta = \left\{ \begin{array}{ll}
           	\eta_{0}  \ ,  & J < J_{\rm EH} \ , \\[3mm]
		 \eta_{0}  \pr{\dfrac{J}{J_{\rm EH}}}^{1/\epsilon ~ - 1}  \ , &   J_{\rm EH} < J < 1000^{\epsilon/(1-\epsilon)} J_{\rm EH}   \ , \\[3mm]
		1000  \eta_{0}  \ ,  &  1000^{\epsilon/(1-\epsilon)} J_{\rm EH} < J     \ ,    
\end{array} \right. 	
\end{equation}
where $\eta_{0}$ is the initial resistivity, $\epsilon$ is a constant value sufficiently less than unity, and $J_{\rm EH}$ is the current density at which the electron heating sets in.
In this paper, we take $\epsilon$ to be 0.1, and $J_{\rm EH}$ to be the arbitrary parameter.
Here, we assume that the ion resistivity is higher than the electron resistivity by a factor of 1000.

At $E_{\rm EH} < E' \lesssim 1000 E_{\rm EH}$, current density is approximately equal to $J_{\rm EH}$ in this model.
Therefore, $J_{\rm EH}$ also approximately corresponds to the saturated current density.
The difference between the saturated current density and $J_{\rm EH}$ is at most smaller than a factor of two.

\subsection{Simulation Settings}

We use a shearing box with a uniform shear flow with the background azimuthal velocity of $- 1.5\Omega x$.
The simulation box sizes in the radial, azimuthal, and vertical direction are $H$, $2\pi H$,  and $H$, respectively, where $H$ is the gas scale height, $c_{s} / \Omega$.
We impose the shearing periodic boundary condition for $x$ and the periodic boundary condition for $y$ and $z$.

We take the computational units of length, time, and density to be, respectively, $H$, $\Omega^{-1}$, and the initial gas density $ \rho_{0}$. 
Therefore, the unit of velocity is $c_{\rm s}$, and the unit of pressure is the initial gas pressure $P_{0} = \rho_{0} c_{\rm s}^{2} $.
The unit of magnetic field strength is 
\begin{equation}\label{eq:}
	B_{\rm u} = \sqrt{4 \pi P_{0} }  \ .
\end{equation}
We take the unit of current density to be
\begin{equation}\label{eq:def-ju}
	J_{\rm u} = \frac{c}{4 \pi} \frac{B_{\rm u}}{H} \ .
\end{equation}
We nondimensionalize the Ohm's law $E=(4 \pi \eta /c^{2}) J$ as $E/E_{\rm u} =  (\eta/\eta_{\rm u})  (J/J_{\rm u})$  \ ,
where
\begin{equation}\label{eq:}
	E _{\rm u} =  \frac{4 \pi \eta_{\rm u} J_{\rm u}}{c^{2}} = \ \frac{c_{\rm s}}{c} B_{\rm u}
\end{equation}
and
\begin{equation}
\eta_{\rm u} =H^{2} \Omega = H c_{\rm s} \ .
\end{equation}

The initial vertical magnetic field is uniform and its strength is 
\begin{equation}\label{eq:}
	B_{z0} =  \sqrt{2} \beta_{0}^{-1/2}  B _{\rm u} \ , 
\end{equation}
where
\begin{equation}\label{eq:}
	\beta_{0} = \frac{8 \pi  P_{0} }{B_{z0}^{2}} 
\end{equation}
is the initial plasma beta.
We consider the situation where MRI would be fully active if electron heating were absent. The activity of MRI is determined by the Elsasser number \citep[e.g.,][]{Sano1999Magnetorotation},
\begin{equation}\label{eq:lamz}
	\Lambda_{z} = \frac{\vAz^{2} }{\eta \Omega} \ ,
\end{equation}
where 
\begin{equation}
\vAz = \frac{B_{z}}{\sqrt{4 \pi \rho}}
\end{equation}
is the \Alfven velocity of the vertical magnetic field.
MRI is fully active when $\Lambda_{z} \gg 1$, while the resistivity suppresses the most unstable MRI mode when $\Lambda_{z} \ll 1$.
We choose the value of $\eta_{0}$ so that the Elsasser number in the initial state $\Lambda_0$ is equal to 10. 
$\Lambda_{0}$ is expressed as $\Lambda_0 = v_{\rm A0}^2/\eta_0 \Omega$,
where $v_{\rm A0}$ is the \Alfven velocity of initial state, $v_{\rm A0} = B_{z0}^{2}/\sqrt{4\pi \rho_{0}}$.
For this value of $\Lambda_0$, the Elsasser number in the final saturated state also satisfies $\Lambda_{z} \gg 1$ as long as electron heating is neglected ($\eta = \eta_0$ for all $E'$) because we generally have $\vAz > \vAzero$. 
In order to investigate dependence on the critical current density $J_{\rm EH}$, we take $J_{\rm EH}$ to be less than $10 J_{\rm u}$ which approximately corresponds to the maximum current density, at which current density is saturated in fully developed MRI turbulence \citep{Muranushi2012Interdependence}.
We give random perturbations of pressure $\delta P$ and velocity $\delta \bm{v}$ whose the maximum amplitude are $\delta P/P_{0} = 5\times 10^{-5}$ and $ |\delta  \bm{v} |/c_{\rm s} = 2 \times 10^{-5} $, respectively.
The amplitudes are taken to be so small that they never exceed the amplitudes of the perturbations left after electron heating suppresses MRI turbulence. 
We also take into account a small viscosity which is effective to damp initial perturbations.

The numerical resolution is taken to be 64, $64/\pi$, and 64 grids per $H$ in the $x$, $y$, and $z$ directions, respectively. 
In order to properly resolve the MRI turbulence, we take the vertical grid spacing $\Delta z$ to be much smaller than the most unstable wavelength $\lambda_{\rm MRI}$ \citep{Noble2010Dependence-of-I}.
In our fiducial model, $\lambda_{\rm MRI}/\Delta z \approx 20$--$120$ in the final state.
In order to resolve MRI, $\lambda_{\rm MRI}/\Delta z \gtrsim 6$ is required \citep{Sano2004Angular-Momentu}.
Our resolution satisfies this requirement. 
The Courant-Friedrichs-Lewy number of 0.4 is used.  

\begin{table*}[t]
\begin{center}
\caption{Summary of results. }
\label{tbl:sim-resu}\smallskip
 
\footnotesize
 \centering
\begin{tabular}{llll | llllll}
\hline \hline
 Label  &$J_{\rm EH}$ & $\beta$ &$\Lambda_{0}$    &$\aved{ B^{2} }/(8\pi P_{0}) $                                &$ \alpha_{\rm M}  $                            &$\alpha_{\rm R} $                         &$\aved{J}/J_{\rm u}$       \\\hline 
EH0001        &   0.001  &   $10^{4}$  &      10    &$1.00 \times 10^{-4}$   &$7.21 \times 10^{-9}$   &$9.00 \times 10^{-11}$   &$1.75 \times 10^{-3}$\\
EH0003        &   0.003  &   $10^{4}$  &      10    &$1.00 \times 10^{-4}$   &$6.49 \times 10^{-8}$   &$8.10 \times 10^{-10}$   &$5.25 \times 10^{-3}$\\
EH001         &    0.01  &   $10^{4}$  &      10    &$1.03 \times 10^{-4}$   &$7.21 \times 10^{-7}$   &$9.00 \times 10^{-9}$   &$1.75 \times 10^{-2}$\\
EH003         &    0.03  &   $10^{4}$  &      10    &$1.30 \times 10^{-4}$   &$6.49 \times 10^{-6}$   &$8.10 \times 10^{-8}$   &$5.25 \times 10^{-2}$\\
EH01          &     0.1  &   $10^{4}$  &      10    &$4.15 \times 10^{-4}$   &$6.47 \times 10^{-5}$   &$3.37 \times 10^{-4}$   &$1.71 \times 10^{-1}$\\
EH03          &     0.3  &   $10^{4}$  &      10    &$1.57 \times 10^{-3}$   &$3.08 \times 10^{-4}$   &$9.73 \times 10^{-4}$   &$4.62 \times 10^{-1}$\\
EH1           &       1  &   $10^{4}$  &      10    &$5.48 \times 10^{-3}$   &$1.53 \times 10^{-3}$   &$1.98 \times 10^{-3}$   &$1.39$\\
EH3           &       3  &   $10^{4}$  &      10    &$1.17 \times 10^{-2}$   &$4.79 \times 10^{-3}$   &$2.95 \times 10^{-3}$   &$3.53$\\
EH10          &      10  &   $10^{4}$  &      10    &$3.75 \times 10^{-2}$   &$1.69 \times 10^{-2}$   &$6.15 \times 10^{-3}$   &$8.23$\\
noEH          &$\infty$  &     $10^{4}$  &    10    &$7.15 \times 10^{-2}$   &$3.15 \times 10^{-2}$   &$9.67 \times 10^{-3}$   &$1.25 \times 10^{1}$\\
\hline
B3-EH         &   0.003  &   $10^{3}$  &      10    &$1.00 \times 10^{-3}$   &$1.41 \times 10^{-7}$   &$1.37 \times 10^{-8}$   &$4.60 \times 10^{-3}$\\
B3-EH         &    0.03  &   $10^{3}$  &      10    &$1.02 \times 10^{-3}$   &$1.41 \times 10^{-5}$   &$1.37 \times 10^{-6}$   &$4.60 \times 10^{-2}$\\
B3-EH         &     0.3  &   $10^{3}$  &      10    &$2.15 \times 10^{-3}$   &$5.99 \times 10^{-4}$   &$3.34 \times 10^{-4}$   &$4.22 \times 10^{-1}$\\
B3-EH         &       3  &   $10^{3}$  &      10    &$4.41 \times 10^{-2}$   &$2.52 \times 10^{-2}$   &$1.08 \times 10^{-2}$   &$3.56$\\
B3-noEH       &$\infty$  &     $10^{3}$  &    10    &$2.22 \times 10^{-1}$   &$1.09 \times 10^{-1}$   &$2.79 \times 10^{-2}$   &$1.55 \times 10^{1}$\\
B5-EH0003     &   0.003  &   $10^{5}$  &      10    &$1.04 \times 10^{-5}$   &$2.70 \times 10^{-8}$   &$4.27 \times 10^{-11}$   &$5.98 \times 10^{-3}$\\
B5-EH003      &    0.03  &   $10^{5}$  &      10    &$4.90 \times 10^{-5}$   &$2.69 \times 10^{-6}$   &$4.29 \times 10^{-9}$   &$5.97 \times 10^{-2}$\\
B5-EH03       &     0.3  &   $10^{5}$  &      10    &$1.66 \times 10^{-3}$   &$1.19 \times 10^{-4}$   &$2.91 \times 10^{-4}$   &$5.25 \times 10^{-1}$\\
B5-EH3        &       3  &   $10^{5}$  &      10    &$4.07 \times 10^{-3}$   &$1.66 \times 10^{-3}$   &$1.35 \times 10^{-3}$   &$3.49$\\
B5-noEH       &$\infty $ &     $10^{5}$  &    10    &$2.56 \times 10^{-2}$   &$1.17 \times 10^{-2}$   &$4.09 \times 10^{-3}$   &$9.34$\\
\hline
L1-EH0003     &   0.003  &   $10^{4}$  &       1    &$1.00 \times 10^{-4}$   &$3.89 \times 10^{-8}$   &$4.86 \times 10^{-10}$   &$4.07 \times 10^{-3}$\\
L1-EH003      &    0.03  &   $10^{4}$  &       1    &$1.18 \times 10^{-4}$   &$3.89 \times 10^{-6}$   &$4.86 \times 10^{-8}$   &$4.07 \times 10^{-2}$\\
L1-EH03       &     0.3  &   $10^{4}$  &       1    &$8.58 \times 10^{-4}$   &$1.64 \times 10^{-4}$   &$1.04 \times 10^{-3}$   &$3.50 \times 10^{-1}$\\
L1-EH3        &       3  &   $10^{4}$  &       1    &$1.08 \times 10^{-2}$   &$4.07 \times 10^{-3}$   &$3.16 \times 10^{-3}$   &$2.78$\\
L1-noEH       &$\infty$  &     $10^{4}$  &     1    &$3.64 \times 10^{-2}$   &$1.67 \times 10^{-2}$   &$5.82 \times 10^{-3}$   &$8.20$\\
L30-EH0003    &   0.003  &   $10^{4}$  &      30    &$1.00 \times 10^{-4}$   &$8.28 \times 10^{-8}$   &$1.03 \times 10^{-9}$   &$5.94 \times 10^{-3}$\\
L30-EH003     &    0.03  &   $10^{4}$  &      30    &$1.39 \times 10^{-4}$   &$8.28 \times 10^{-6}$   &$1.03 \times 10^{-7}$   &$5.94 \times 10^{-2}$\\
L30-EH03      &     0.3  &   $10^{4}$  &      30    &$1.84 \times 10^{-3}$   &$3.80 \times 10^{-4}$   &$1.32 \times 10^{-3}$   &$5.22 \times 10^{-1}$\\
L30-EH3       &       3  &   $10^{4}$  &      30    &$1.36 \times 10^{-2}$   &$5.68 \times 10^{-3}$   &$2.83 \times 10^{-3}$   &$3.93$\\
L30-noEH      &$\infty$  &     $10^{4}$  &    30    &$7.28 \times 10^{-2}$   &$3.26 \times 10^{-2}$   &$1.01 \times 10^{-2}$   &$1.28 \times 10^{1}$\\

\hline 
\end{tabular}
\end{center}
\end{table*}

\subsection{Initial Conditions}
We take $\beta_{0} = 10^{4}$ and $\Lambda_{0} = 10$ as the fiducial parameters.
For this set of $\beta_0$ and $\Lambda_0$, we 
consider 10 different values of $J_{\rm EH}$: $J_{\rm  EH}/J_{\rm u} =  1\times10^{-3}, 3\times10^{-3}, 1\times10^{-2}, 3\times10^{-2}, 1\times10^{-1}, 3\times10^{-1}, 1, 3, 10$ and $\infty$, where $J_{\rm EH}/J_{\rm u} = \infty$ corresponds to the case without electron heating.
We also perform simulations with different values of $\beta_{0}$ and $\Lambda_{0}$ to see the dependence on these parameters. 
We take $\beta_{0}$ as $\beta_{0} = 10^{3} , 10^{4} , 10^{5}$ and $\Lambda_{0} $ as $\Lambda_{0} = 30 , 10 , 0.3$, with $J_{\rm EH} =  0.003, 0.03, 0.3, 3,  \infty$ for each set of $\beta_{0}$ and $\Lambda_{0}$. 
We use these results for checking accuracy of the analytic $\alpha_{\rm M}$--$J_{\rm EH}$ relation presented in Section \ref{sec:der-pred-form}.

\section{Simulation Results}\label{sec:sim-results}

Table \ref{tbl:sim-resu} summarizes the parameter sets explored in this study.
We express the volume-averaged quantities as $\ave{...}$  and the time- and volume-averaged quantities as $\ave{\ave{...}}$.
The volume averages are calculated over the entire simulation box, and the time averages are calculated from $100$ to $150$  in units of the orbital period $2 \pi /\Omega$. 
The range of time integration is taken so that the final saturated state dominates the average.

The most important quantity obtained from the simulations is the accretion stress, which controls the disk evolution. 
The accretion stress can be characterized in terms of the Shakura-Sunyaev alpha parameter $\alpha$, 
which is defined as the time- and volume-averaged accretion stress divided by the time- and volume-averaged pressure, which is equal to $P_{0}$ for an isothermal gas, 
\begin{equation}
\alpha =  \alpha_{\rm R} +  \alpha_{\rm M}   = \frac{ \aved{\rho v_{x} \delta v_{y} } }{ P_{0} } + \frac{ \aved{-B_{x} B_{y} }}{ 4 \pi P_{0}},
\end{equation}
where we express $ \aved{\rho v_{x} \delta v_{y} }/ P_{0}$  and $\aved{-B_{x} B_{y} }/ (4 \pi P_{0})$ as, respectively, $\alpha_{\rm R}$ and $\alpha_{\rm M}$.

\begin{figure*}[t]
\centering
    \includegraphics[width=0.4 \hsize,clip]{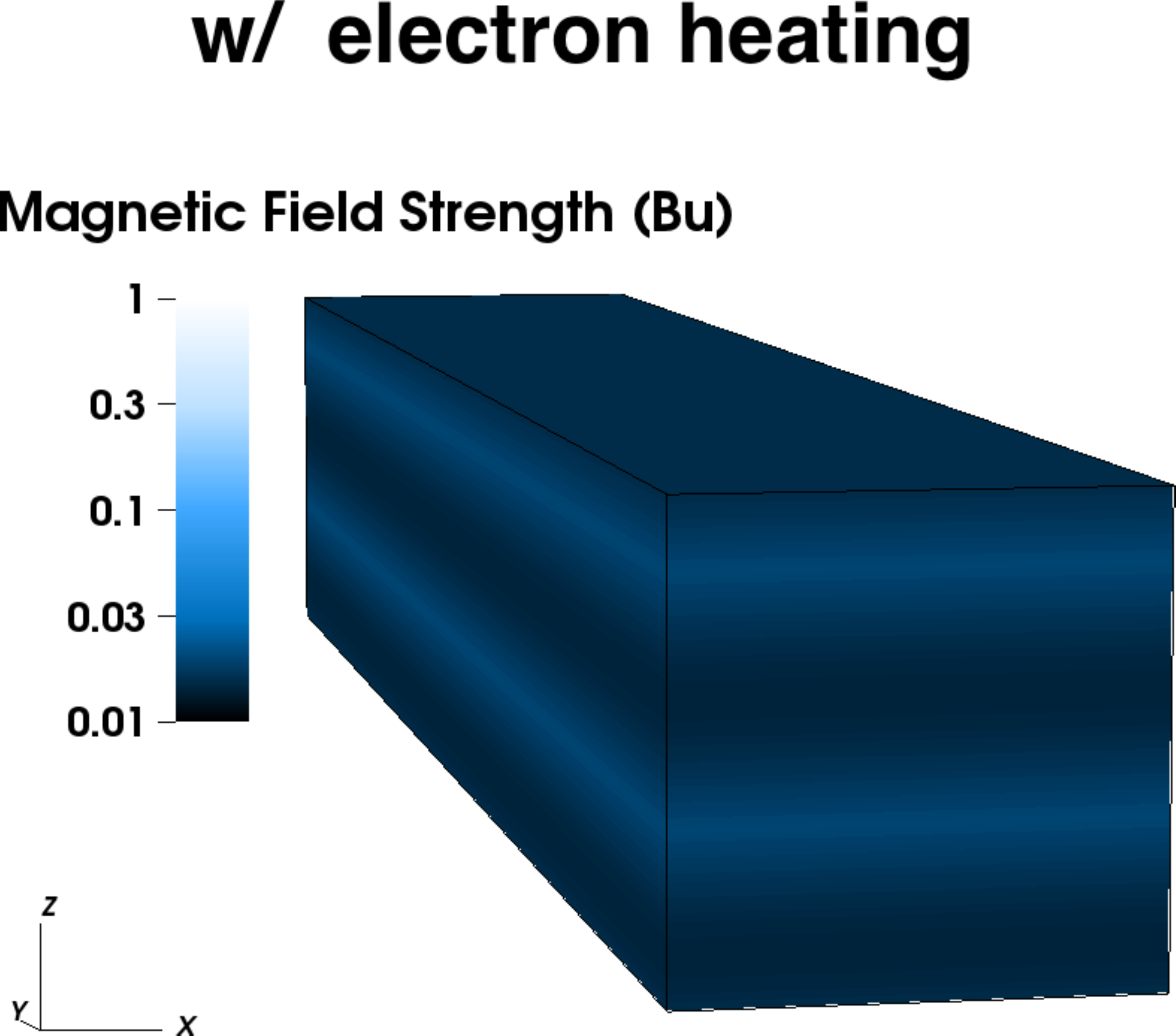} \hspace{1cm}
  \includegraphics[width=0.4 \hsize,clip]{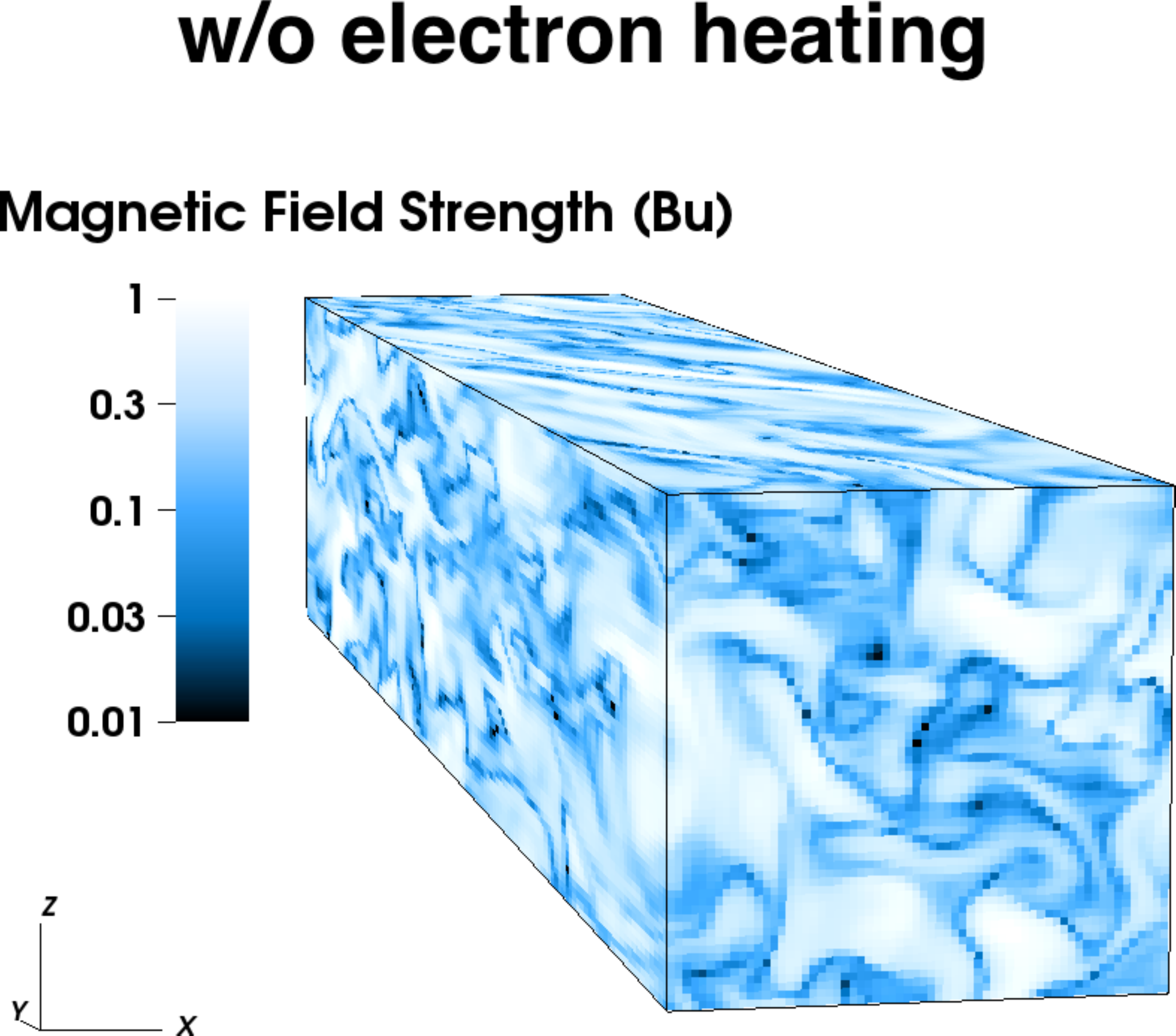} 
\caption{Snapshot of magnetic field strength $|\bm{B}|/B_{\rm u}$ at 60 orbits for $J_{\rm EH}/J_{\rm u} = 0.03$ (left) and for the case without electron heating $J_{\rm EH}/J_{\rm u} = \infty$ (right).}
 \label{fig:outline}
\end{figure*}

\begin{figure*}[t]
\centering 
 \includegraphics[width=0.8\hsize,clip]{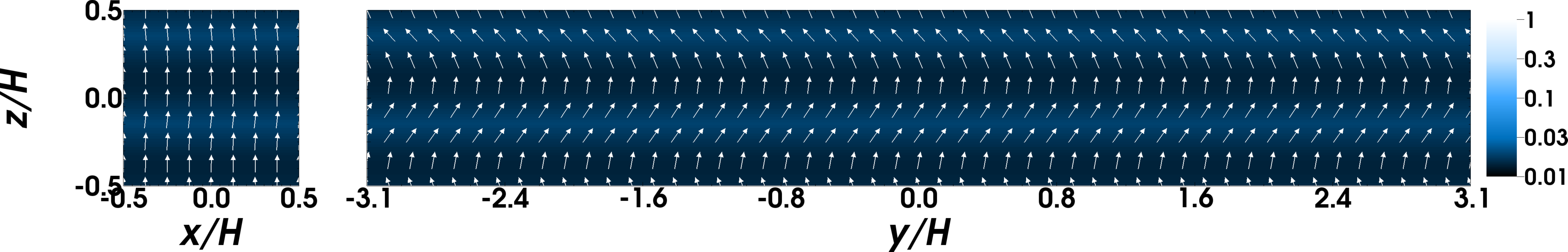}\vspace{2mm}
 \includegraphics[width=0.8\hsize,clip]{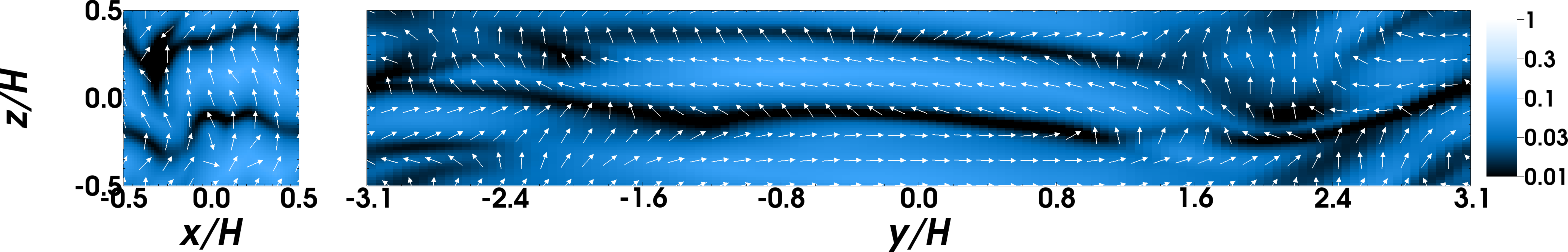}\vspace{2mm}
 \includegraphics[width=0.8\hsize,clip]{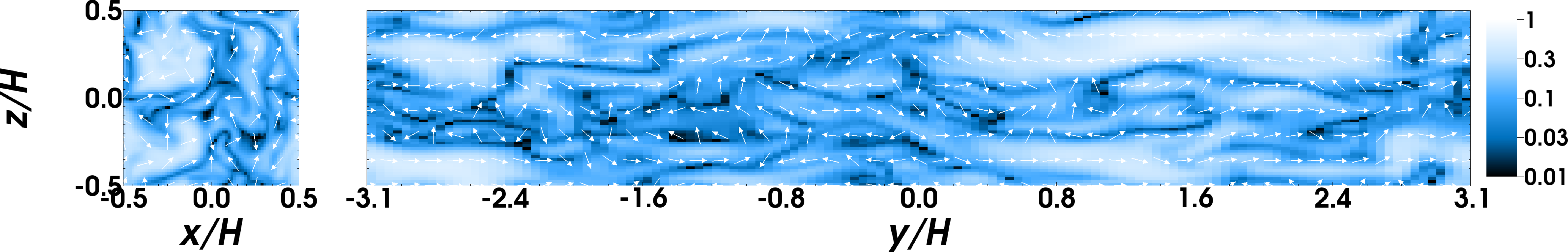}\vspace{2mm}
 \includegraphics[width=0.8\hsize,clip]{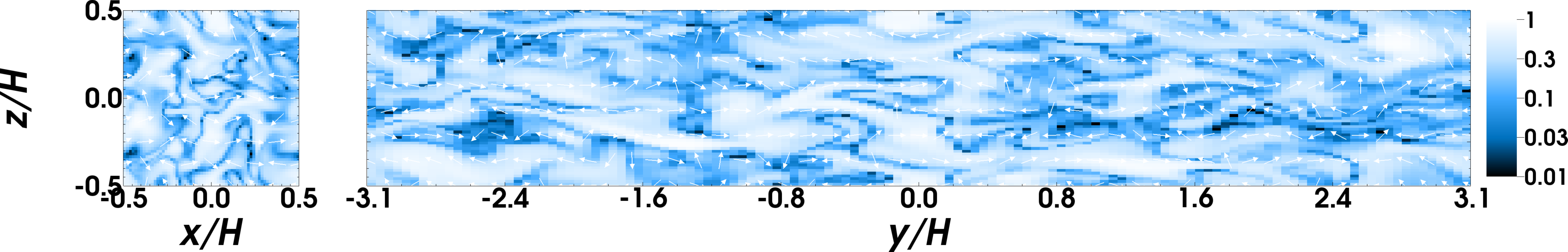}
\caption{Slices in the $x$--$z$ plane at $y=0$ and in the $y$--$z$ at $x=0$ of the magnetic field strength $|\bm{B}|/B_{\rm u}$ (color) and direction of the magnetic field (arrows) at 60 orbits for $J_{\rm EH}/J_{\rm u} = 0.03$, 0.3, 3, and $\infty$, from top to bottom.}
 \label{fig:sim-2d-xz-yz}
\end{figure*}

\subsection{The Fiducial Case}\label{ssec:rep-res}

Figure \ref{fig:outline} shows the saturated state ($t= 60$ orbits) observed in our fiducial simulations with $J_{\rm EH}/J_{\rm u} = 0.03$.
The saturated state for the case without electron heating $J_{\rm EH}/J_{\rm u} = \infty$ is also shown for comparison. 
We also show the crosscuts of the saturated state on the $x$--$z$ and $y$--$z$ planes for $J_{\rm EH} = 0.03, 0.3, 3,$ and $\infty$ in Figure \ref{fig:sim-2d-xz-yz} .
We find that a laminar flow with an ordered magnetic field dominates the saturated state for $J_{\rm EH}/J_{\rm u} = 0.03$, 
whereas the turbulent magnetic fields are generated in the case without electron heating.
Comparing these two case, we confirm that electron heating suppress turbulent motion that is characteristic of MRI. 
Moreover, the magnetic field strength $|\bm{B}|$ is also largely suppressed for the laminar case.

In Figure \ref{fig:sim-2d-xz-yz}, we see that the azimuthal magnetic fields for $J_{\rm EH}/J_{\rm u} = 0.03$ are sinusoidal in the vertical direction, with a wavelength as large as the vertical box size. 
In the presence of electron heating, the perturbations on small scales stop growing due to the increased resistivity, while perturbations on larger scales grow.
For this reason, the magnetic field inside the box tends to be dominated by the component whose wavelength is equal to the box size.
We also see that small structure of magnetic fields appears with increasing $J_{\rm EH}$.
This too can be understood by the fact that the resistivity increased by the electron heating suppresses the perturbations on small scale.

\begin{figure}[t]
\centering
\includegraphics[width=\hsize]{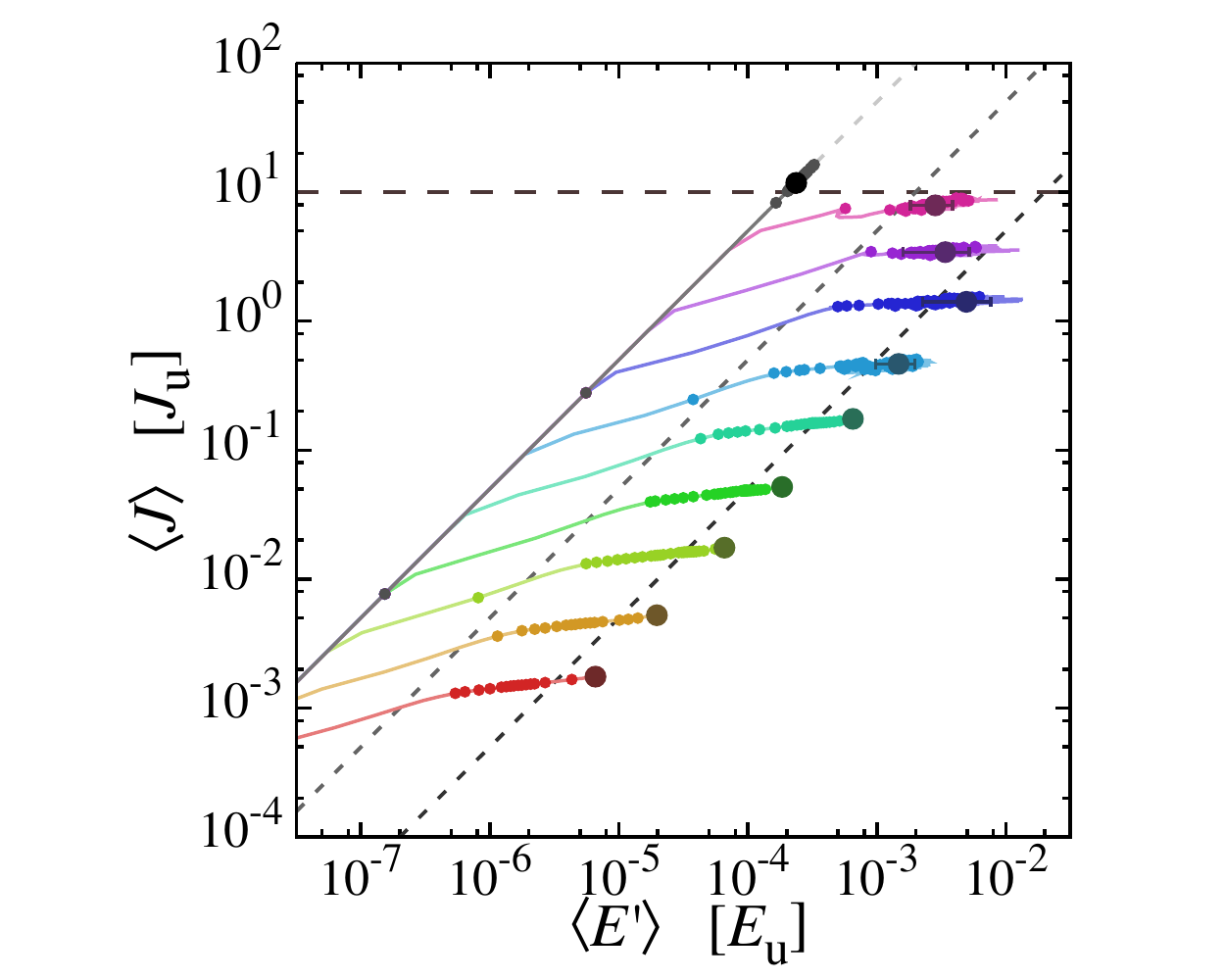} 
\caption{Evolution tracks of the volume-averaged electric field strength $\ave{E'}$ ($x$-axis) and current density $\ave{J}$ ($y$-axis) mapped in the $J$--$E'$ plane.
Curves of different colors correspond to runs of different values of $J_{\rm EH}$ (from bottom to top,  $J_{\rm EH}/J_{\rm u} = 0.001, 0.003, 0.01, 0.03, 0.1, 0.3, 1, 3, 10$ and $\infty$).
The dashed lines indicate the linear relations $\ave{J} = c^2/(4\pi\eta_0) \ave{E'}$, where 
$\eta_0 = \vAzz^2/\Lambda_0 \Omega$ (see the text after \eqref{eq:lamz}) with $\Lambda_0 =$ 10 (light gray line), 1 (gray line), and 0.1 (dark gray line), respectively.
The small circles are plotted at every one orbit to visualize the rates of change in $\ave{J}$ and $\ave{E'}$.
The dark filled circles indicate the final saturated states.
The horizontal dotted line is $J = 10 J_{\rm u}$, which is the saturated current density in the fully developed MRI turbulence \citep{Muranushi2012Interdependence}.
}
 \label{fig:sim-e-cur}
\end{figure}

In order to see that the resulting $J$ and $E'$ follow the given $J$--$E'$ relation, 
in Figure \ref{fig:sim-e-cur}, we show the evolutionary tracks of the volume-averaged current density $\ave{J}$ and electric field strength $\ave{E'}$ in the $J$--$E'$ plane. 
The current densities initially grow along the line of $\Lambda_{0}=10$ and then branch off the line after they reach $J_{\rm EH}$.
We confirm the resulting $\ave{J}$--$\ave{E'}$ tracks almost go along with $J$--$E'$ relation we give.
We also find that, in the absence of electron heating cases, $\ave{J}$ and $\ave{E'}$ are saturated near the line corresponding to $\Lambda_0 = 0.1$. 

\begin{figure}[t]
\centering
  \includegraphics[width=\figscale\hsize,clip]{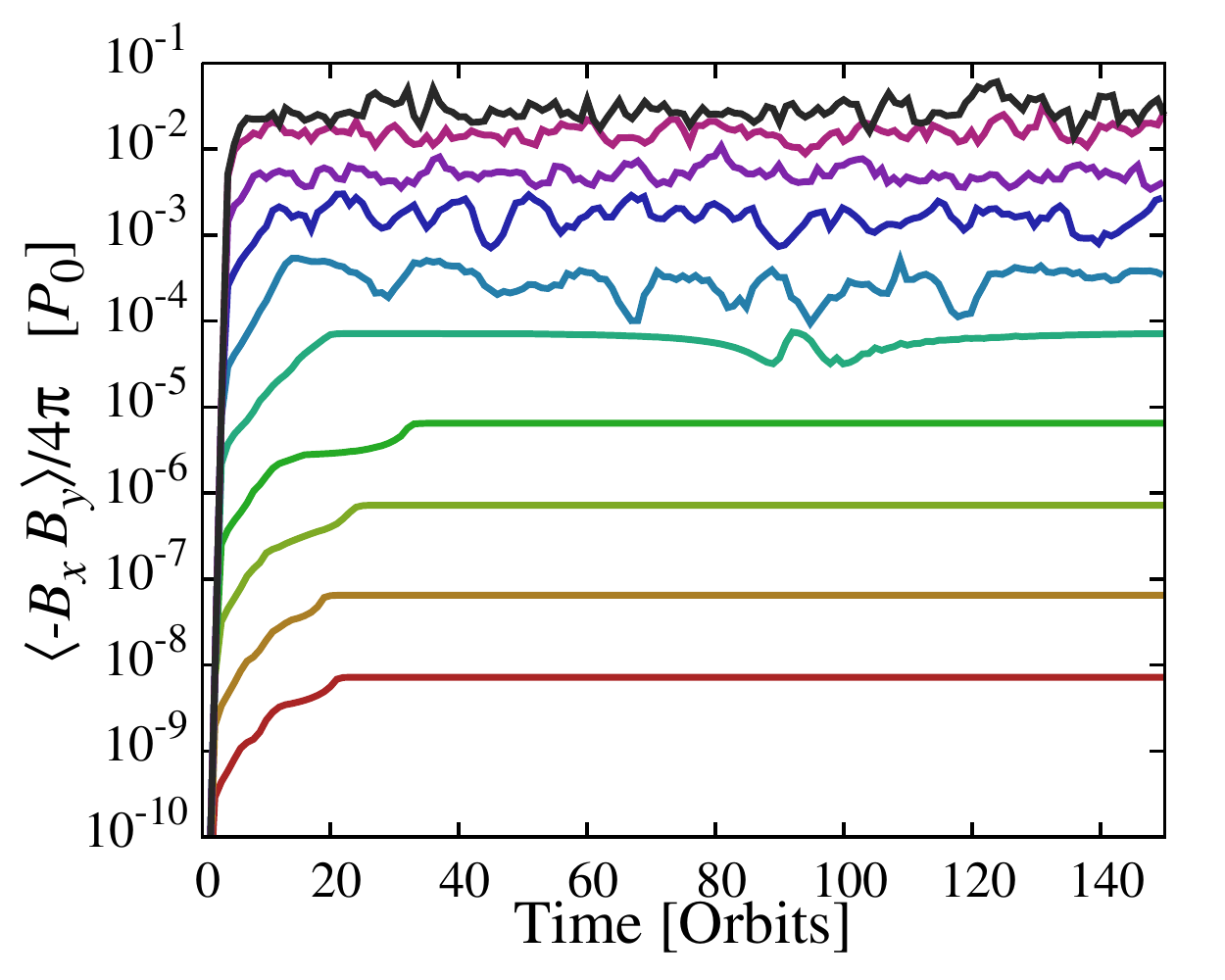} 
\caption{Time evolution of the volume-averaged Maxwell stress for different values of $J_{\rm EH}$ (from bottom to top,  $J_{\rm EH}/J_{\rm u} = 0.001, 0.003, 0.01, 0.03, 0.1, 0.3, 1, 3, 10$ and $\infty$).
 The color scheme is the same as in Figure \ref{fig:sim-e-cur}.
 }
 \label{fig:sim-alp}
\end{figure}

In Figure \ref{fig:sim-alp}, we show the time evolution of the volume-averaged Maxwell stress for different values of $J_{\rm EH}$.
MRI grows linearly in the first few orbits, and then the Maxwell stress becomes saturated in $\sim$ 30 orbits.
We find that the Maxwell stress in the saturated state decreases with decreasing $J_{\rm EH}$, which means that MRI is stabilized by electron heating. 
We also find that the Maxwell stress in the saturated state is fluctuating when $J_{\rm EH}/J_{\rm u} > 0.3$ and is highly stationary when $J_{\rm EH}/J_{\rm u} < 0.1$. 
This suggests that electron heating completely suppresses turbulent motion caused by MRI when $J_{\rm EH} < 0.1 J_{\rm u}$.
We here define the threshold current density as 
\begin{equation}
	J_{\rm lam} = 0.1 J_{\rm u} \ .
\end{equation}
At $J \lesssim J_{\rm lam}$, the saturated state is laminar.

\begin{figure}[t]
	\centering
	\includegraphics[width=\figscale\hsize,clip]{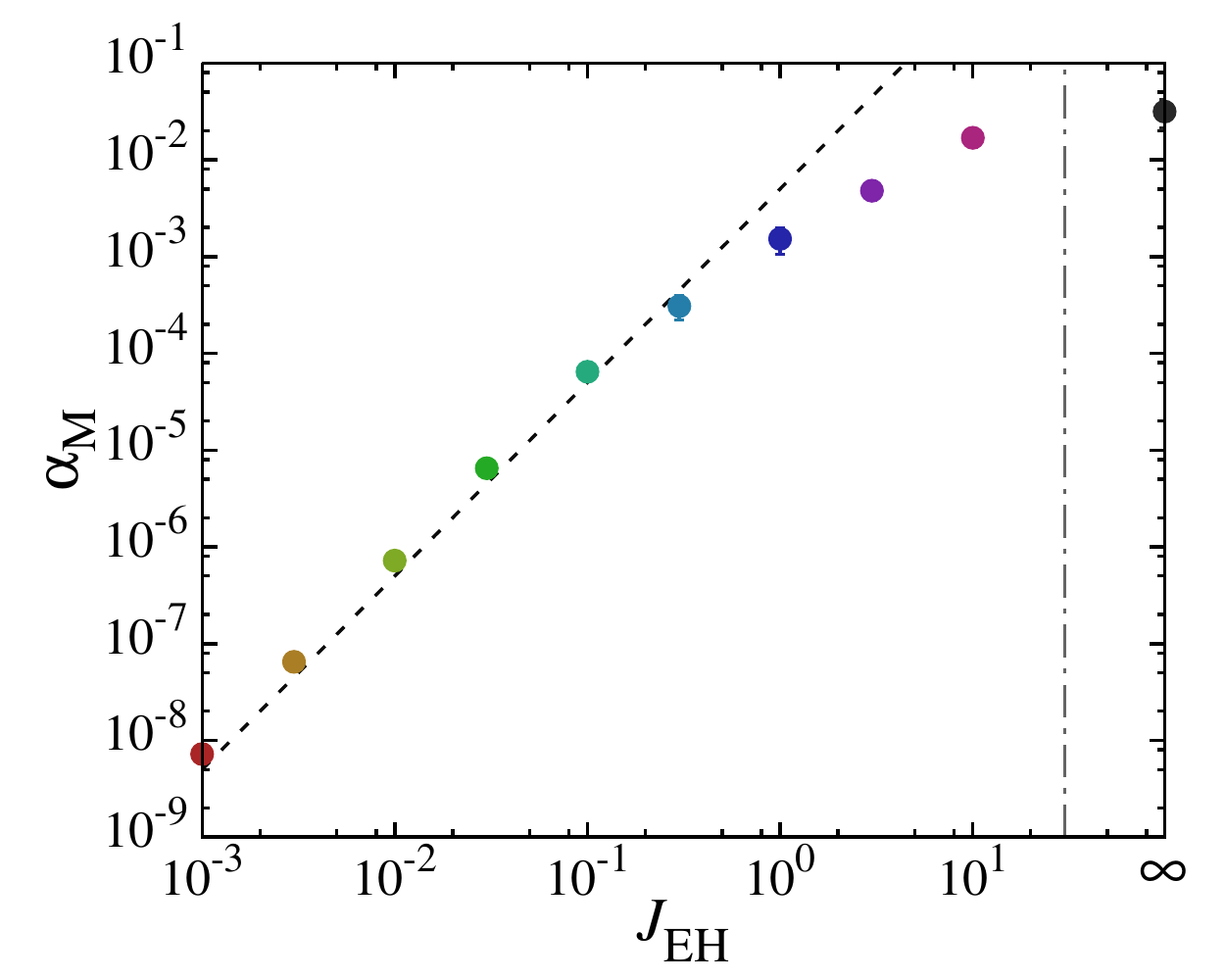} 
	\caption{Time- and volume-averaged Maxwell stress $\alpha_{\rm M}$ as a function of $J_{\rm EH}$ (colored dots).
	The color scheme is the same as in Figure \ref{fig:sim-e-cur}.
	The dashed line shows a quadratic fit for low $J_{\rm EH}$, $ \alpha_{\rm M}  = 0.5 ( J_{\rm EH} /10J_{\rm u})^{2}$.}
	 \label{fig:sim-maxwell} 
\end{figure}

Figure \ref{fig:sim-maxwell} displays $\alpha_{\rm M}$ as a function of $J_{\rm EH}$.
We confirm a positive correlation between $\alpha_{\rm M}$ and $J_{\rm EH}$. 
By fitting a quadratic function to the data, we obtain the empirical formula of the relation, 
\begin{equation}\label{eq:emp-alp}
	\alpha _{\rm M} = 0.5 \pf{J_{\rm EH}}{10 J_{\rm u}}^{2} \ ,
\end{equation}
The dependence on current density, $\alpha_{\rm M}  \propto  J^{2} $, is consistent with a scaling relation obtained by \citetalias{Mori2016Electron-Heatin} (Equation (40) in their paper), although the magnitude in their equation is 50 times smaller than obtained here.
This empirical fit can be used when $J_{\rm EH} $ is less than $J_{\rm lam}$.

\begin{figure}[t]
	\centering
	\includegraphics[width=\figscale\hsize,clip]{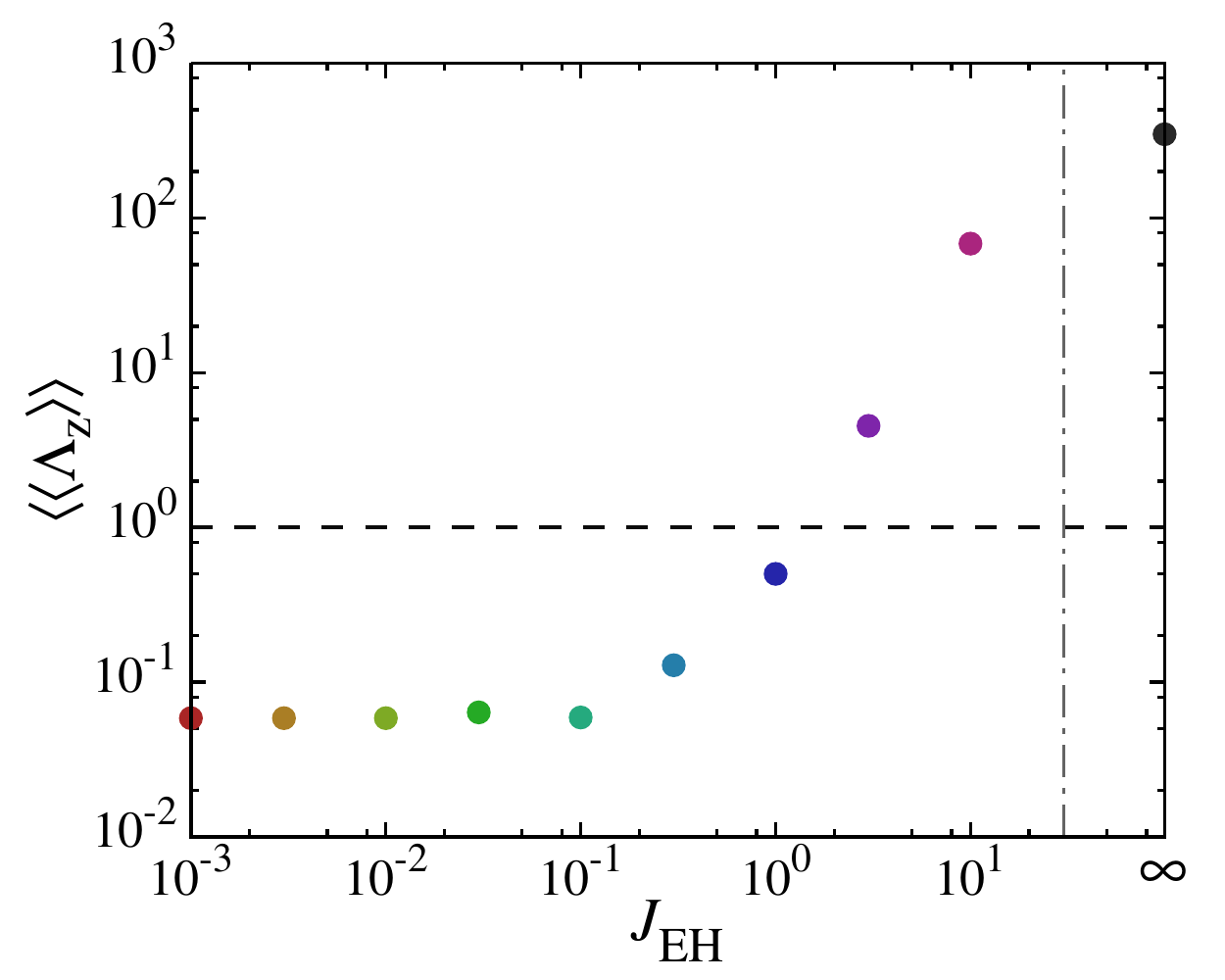}  \vspace{-2mm}
	\caption{Time- and volume-averaged Elsasser number $\aved{\Lambda_{z}}$ as a function of $J_{\rm EH}$. The dashed line shows $\Lambda = 1$. The color scheme is the same as in Figure \ref{fig:sim-e-cur}. 
	}
	\label{fig:sim-els-J}
	\vspace{-0mm}
\end{figure}

According to previous studies \citep[e.g.,][]{Sano2002bThe-Effect-of-t}, $\Lambda_{z}$ expresses the MRI activity.
When the Elsasser number is much higher than unity, MRI can make vigorous magnetic turbulence.
Figure \ref{fig:sim-els-J} shows the volume- and time-averaged Elsasser number $\aved{\Lambda_{z}}$ as a function of $J_{\rm EH}$.
For $1 < J_{\rm EH}/J_{\rm u} < 10$, we see that  although the Elsasser number is higher than unity , $\alpha_{\rm M}$ gradually decreases with decrease of $J_{\rm EH}$ as we see in Figure \ref{fig:sim-maxwell}.
Because the increased resistivity can suppress magnetic fields by the small scale turbulent motion which forms strong current density, 
the electron heating takes place when the MRI turbulence is generated.
We also see that  $\aved{\Lambda_{z}}$ is constant at $J_{\rm EH} < J_{\rm lam}$.
This is because $\eta$ is also constant for $J_{\rm EH} < J_{\rm lam}$ as we see below.

\begin{figure}[t]
\centering
   \includegraphics[width=\figscale\hsize,clip]{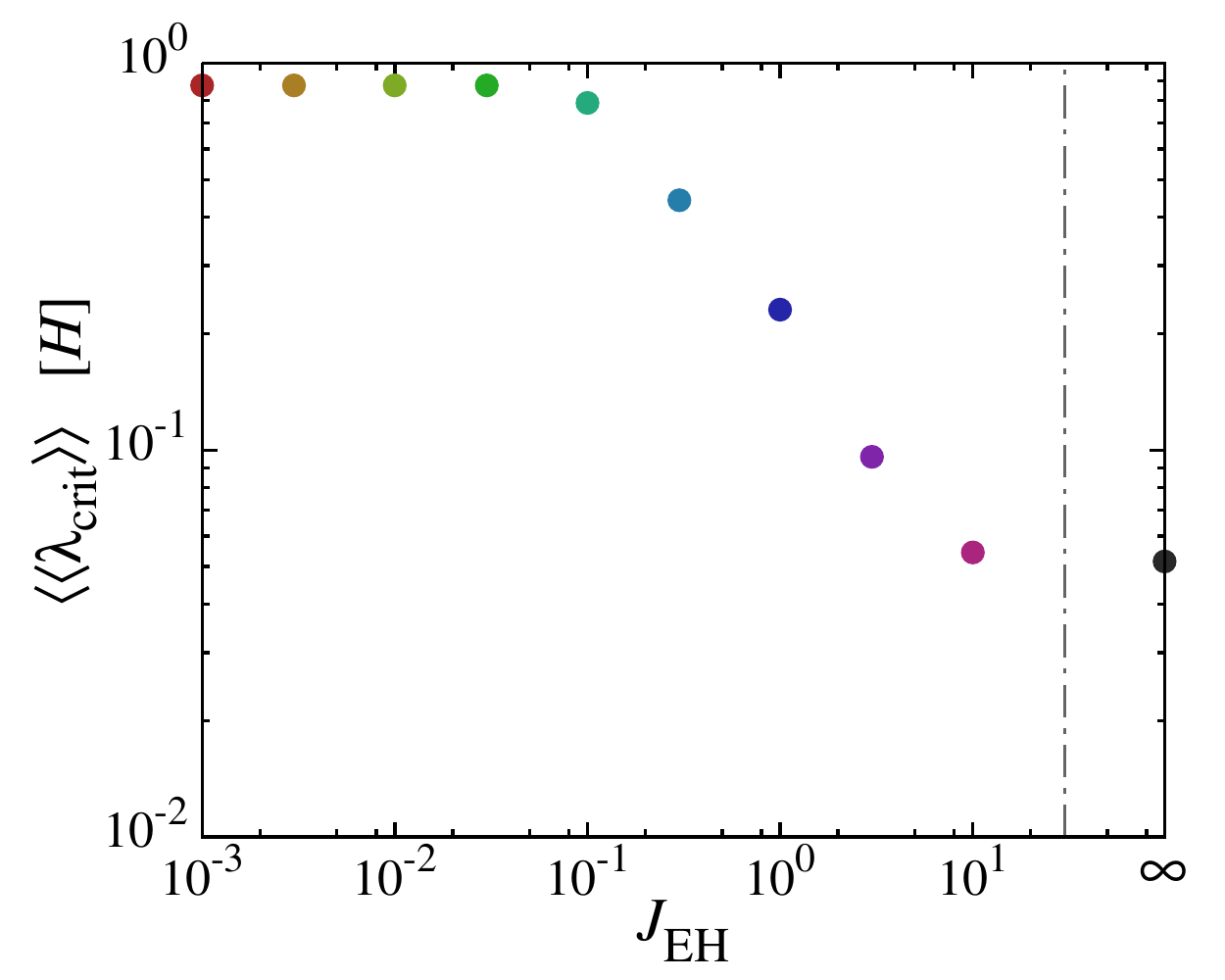}
\caption{Time- and volume-averaged critical wavelength $\aved{\lambda_{\rm crit}}$ as a function of $J_{\rm EH}$. The color scheme is the same as in Figure \ref{fig:sim-e-cur}.}
 \label{fig:sim-muwl-cwl}
\end{figure} 

\begin{figure}[t]
\centering
  \includegraphics[width=\figscale\hsize,clip]{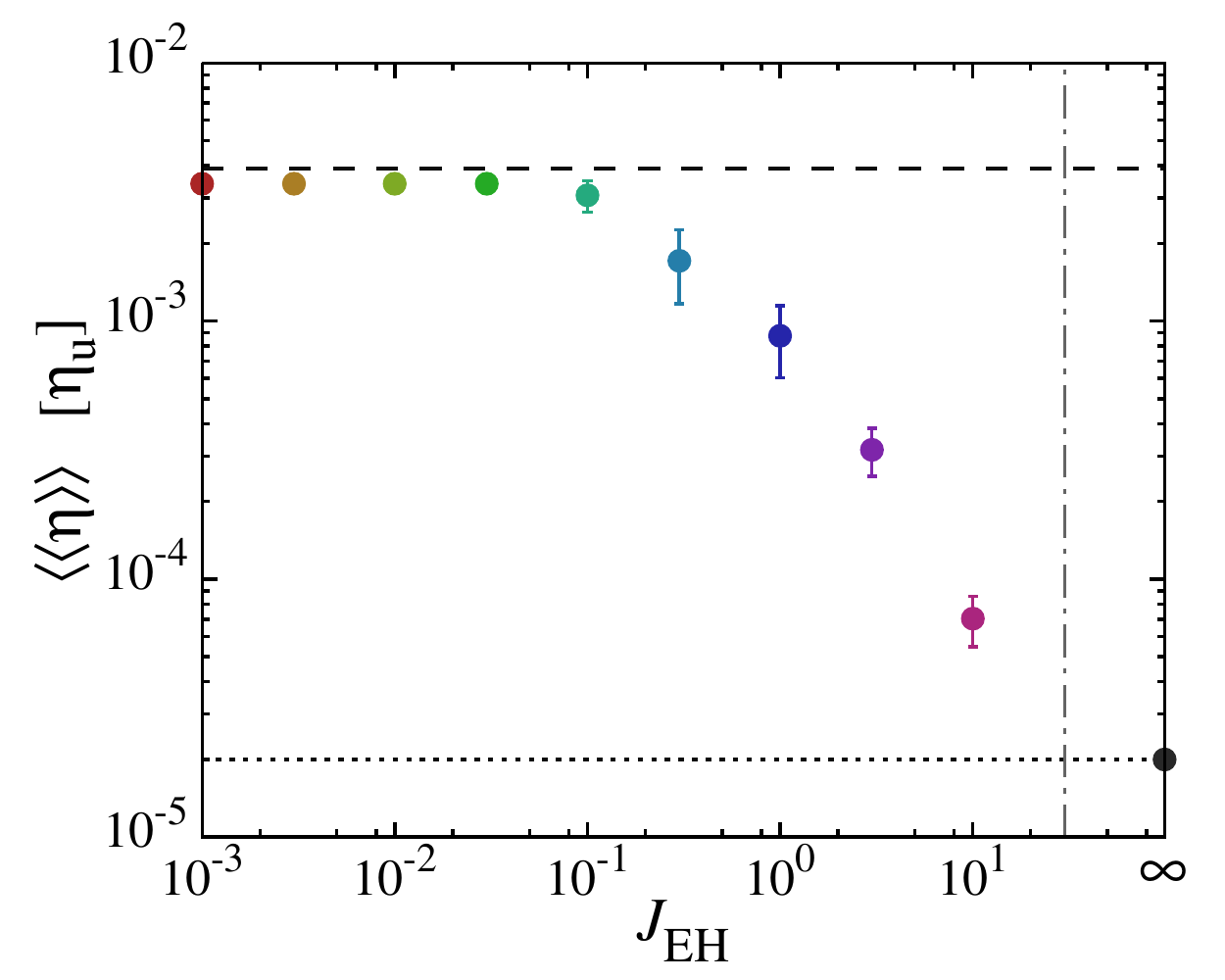} 
\caption{Time- and volume-averaged resistivity $\aved{\eta}$ as a function of $J_{\rm EH}$. The dotted line shows the initial resistivity $\eta_{0}$ and the dashed line shows \eqref{eq:ana-etasat}. The color scheme is the same as in Figure \ref{fig:sim-e-cur}. }
 \label{fig:eta}
\end{figure}

To see why the MRI is quenched in the laminar saturated state,
we show the time- and volume-averaged critical wavelength $\aved{\lambda_{\rm crit}}$ in Figure \ref{fig:sim-muwl-cwl}.
The critical wavelength $\lambda_{\rm crit}$ is the shortest wavelength in unstable MRI mode.
This is obtained from the linearized equation system in \citet{Sano1999Magnetorotation} by assuming growth rate of zero. 
The critical wavelength in both resistive and ideal MHD is written as 
\begin{equation}\label{eq:lam-crit}
	\lambda_{\rm crit} = 2 \pi \frac{1}{\sqrt{3}}  \frac{\vAzz }{\Omega}  \pr{1 + \pf{\vAzz^{2}}{\eta \Omega}^{-2} }^{1/2} \ ,
\end{equation} 
where $\vAzz = B_{z0}/\sqrt{4 \pi \rho_{0}}$. 
We see that the resulting critical wavelength is approximately equal to simulation box size $H$ for low $J_{\rm EH}$.
The MRI growth increases $\eta$, which in turn increases the critical wavelength $\lambda_{\rm crit}$ when $\Lambda \lesssim 1$.
For this reason, the shortest unstable wavelength increases until the wavelength reaches to the box size,
 and eventually all MRI unstable modes die away. 
Note that the final state of this simulation would depend on the vertical box size.

Figure \ref{fig:eta} shows the time- and volume-averaged resistivity $\aved{\eta}$ as a function of $J_{\rm EH}$.
In all simulations but with $J_{\rm EH} = \infty$, the final resistivity is higher than the initial value $\eta_{0}$ (shown by the dotted line).
We see that the saturated resistivity for low $J_{\rm EH}$ is independent of $J_{\rm EH}$.
This value is given by $\lambda_{\rm crit}(\eta) = H$ in the resistive MHD,
\begin{equation}\label{eq:ana-etasat}
	  \frac{ \eta_{\rm lam} }{\eta_{\rm u} } = \frac{2}{ \sqrt{ \beta_{0} 8 \pi^{2} /3 }  } \approx  0.390\times 10^{-2} \pf{\beta_{z0}}{10^{4}}^{-1/2}  \ .
\end{equation}
The resistivity cannot exceed this value because any higher resistivity would stabilize all unstable modes that can fit in the simulation box.
The fact that $\aved{\eta}$ reaches this critical value explains why the laminar saturated state is realized for $J_{\rm EH} < 0.1 J_{\rm u}$.

We see in Figure \ref{fig:sim-alp} that the saturated state for the low $J_{\rm EH}$ is steady.
Although Figures \ref{fig:sim-2d-xz-yz} shows that the wavelength in the final state is equal to the vertical box size, the process to the saturated state has not been shown. 
How is the saturated laminar state determined?
In the presence of electron heating, the resistivity also increases with the unstable mode growing. 
When the increased resistivity reaches the critical resistivity \eqref{eq:ana-etasat}, 
MRI is stabilized since the all unstable mode dies away. 
In this state, if perturbations of magnetic fields grow, then the resistivity is increased and in turn stabilizes the perturbations.
On the other hand, if the perturbation is damped from the equilibrium state, then the resistivity becomes smaller and MRI grows again.
In other word, the saturated laminar state is determined by the balance between the MRI growth by shear and decay by the increased resistivity. 
Therefore, the final state must settle into the stable equilibrium state.

\begin{figure}[t]
\centering
    \includegraphics[width=\figscale\hsize,clip]{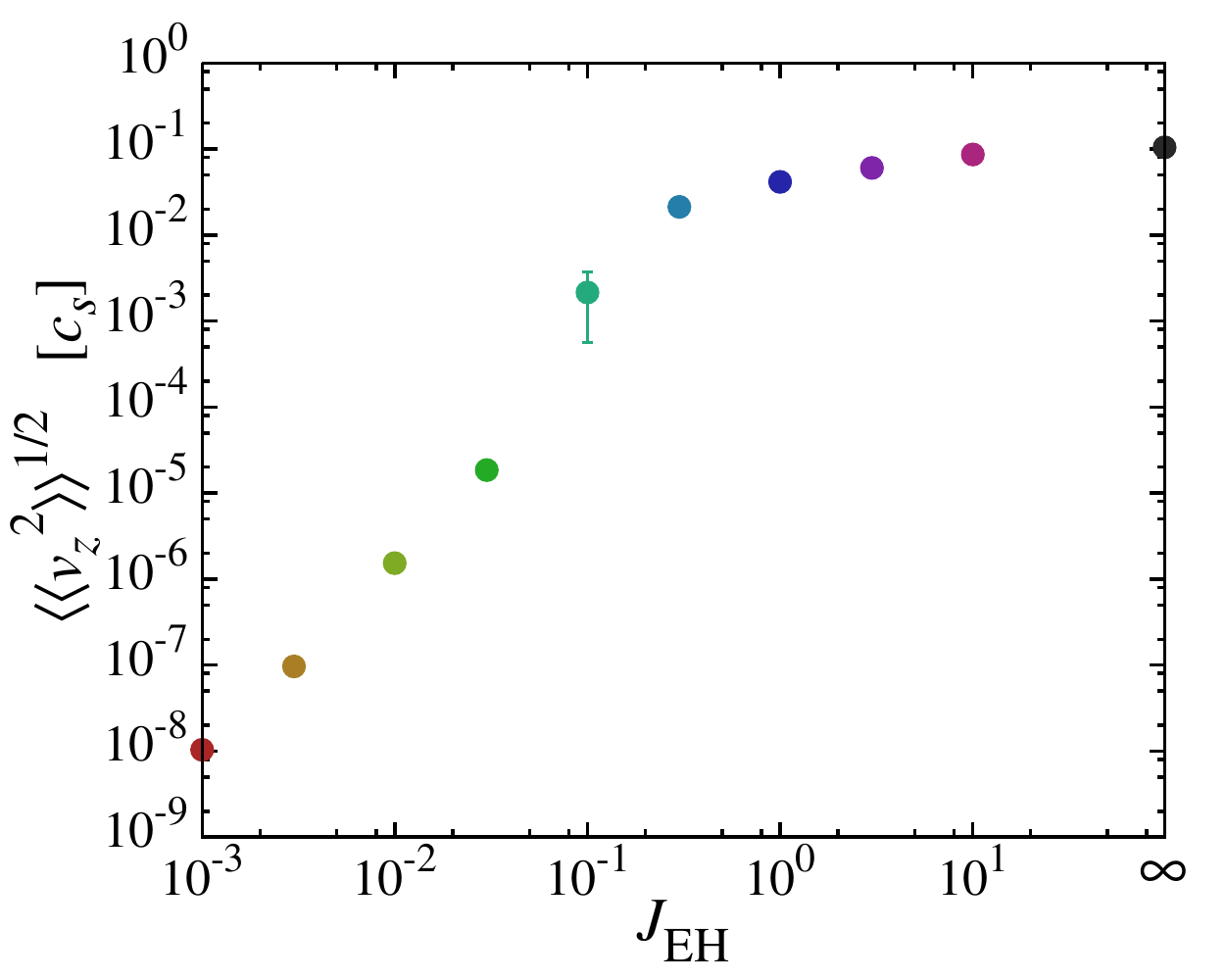} 
\caption{Time- and volume-averaged vertical velocity $\aved{v_{z}^{2}}^{1/2}$ as a function of  $J_{\rm EH}$. The color scheme is the same as in Figure \ref{fig:sim-e-cur}. }
 \label{fig:sim-dif}
\end{figure}

Lastly, in order to see turbulent activity, we plot the root mean square of the vertical velocity $\aved{v_{z}^{2}}^{1/2}$ as a function of $J_{\rm EH}$ in Figure \ref{fig:sim-dif}.
In particular, the vertical velocity of gas is important for dynamics and spatial distribution of dust in protoplanetary disks.
We see that the vertical velocity sharply drops at $J_{\rm EH}\lesssim J_{\rm lam}$, where the saturated state is laminar.
Its implications for turbulent mixing of dust particles are discussed in Section \ref{sec:Summary}.

\section{Derivation of Current--Stress Relation}\label{sec:der-pred-form}  
In this section, we derive a relation between $\alpha_{\rm M}$ and $J_{\rm EH}$ which reproduce our simulation results.
Because $J_{\rm EH}$ can be calculated from disk parameters, this relation may provide a quantitative prediction for accretion stress without MHD simulations, when the saturated state is determined by the electron heating.
For example, this relation would be useful for simplified modeling with disk evolution using $\alpha$ parameter based on MHD simulation with electron heating.
We here neglect contribution of Reynolds stress to accretion stress.
This is because Maxwell stress is generally larger than Reynolds stress according to Table 3. 
In addition, we regard the current density in the saturated state as $J_{\rm EH}$. 

We first derive an analytical expression of the Maxwell stress in the laminar state, $\alpha_{\rm M, lam}$.
To express $\alpha_{\rm M} =  \aved{ - B_{x}B_{y} }/ (4 \pi P_{0})   $ as a function of $J_{\rm EH}$, we estimate $ - B_{x}B_{y}  / 4 \pi P_{0} $ by using the Amp\'ere's equation $ \bm{J} = c/(4 \pi) \nabla \times  \bm{B} $.
We take $\nabla$ to be the typical wavenumber $ \bm{k} $.
We here consider the vertical sinusoidal wave as we see Figure \ref{fig:sim-2d-xz-yz}, and therefore $\bm{k} = k_{z} \bm{{\rm e}_{z}}$ is assumed. 
The $x$-direction component of the current density is described as $	J_{x} \approx  -  ck_{z} B_{y}/4 \pi    $, and thereby $B_{y}$ is written as 
\begin{eqnarray}\label{eq:jx-jy}
B_{y}  \approx - \frac{4 \pi}{c k_{z}} J_{x} \ .
\end{eqnarray}
According to Figure \ref{fig:sim-muwl-cwl}, the critical wavelength in the laminar case is the vertical box size, $H$.
Thus, we here assume that the vertical wavenumber in the saturated state is 
\begin{equation}\label{eq:kz-crit}
	k_{z, {\rm crit}} = \frac{2 \pi}{H} \ .
\end{equation}

Using \eqref{eq:jx-jy} and \eqref{eq:kz-crit}, we express $- B_{x}B_{y} /4 \pi P_{0} $ as 
\begin{equation}\label{eq:bxby-2}
	- \frac{B_{x}B_{y} }{4 \pi P_{0} }  \approx  - \frac{100}{4\pi^{2}}     \pf{B_{x}  }{B_{y}}    \pf{ J }  { 10 J_{\rm u}}^{2}  \ ,
\end{equation}
where the current densities are normalized by typical current density of fully developed turbulence, $\approx 10 J_{\rm u}$, and we assume that $J_{x} \approx J$ because $J_{x}$ dominates the total current density $J$.

The relationship between $B_{x}$ and $B_{y}$ is given from the linearized equation system, Equations (10) and (12) in \citet{Sano1999Magnetorotation},
\begin{equation}\label{eq:ana-bx}
	B_{x}        =  - \frac{ 2 \vAzz^{2}}{\eta_{\rm lam} \Omega} B_{y}  \ .
\end{equation}
where $\eta_{\rm lam}$ is the resistivity in the laminar case, and we use the fact that the saturated state is steady and resistivity is spatially uniform. 
Thus, we give $B_{x} / B_{y}$ in the laminar state as
\begin{equation}\label{eq:bx-by}
	\frac{B_{x}}{B_{y}}   =   - \frac{4}{\beta_{0}} \frac{\eta_{\rm u}}{\eta_{\rm lam}}   \ .
\end{equation}
The saturated resistivity $\eta_{\rm lam}$ is given by \eqref{eq:ana-etasat}.

Using \eqref{eq:bx-by} and \eqref{eq:ana-etasat} to \eqref{eq:bxby-2} in the saturated state, we obtain $\alpha_{\rm M, lam}$ as
\begin{equation}\label{eq:alp-lam}
	\alpha_{\rm M,lam} = 0.25 \pf{ \beta_{0}}{10^{4} }^{-1/2}  \pf{  J_{\rm EH}  }{ 10 J_{\rm u} }^{2} \ ,
\end{equation}
where we assume $J$ to be equal to $J_{\rm EH}$.
\eqref{eq:alp-lam} approximately equals to the fit in Figure \ref{fig:sim-maxwell}.
The difference of the coefficients between \eqref{eq:alp-lam} and the fit comes from the difference between the saturated current density and $J_{\rm EH}$.
Although \eqref{eq:alp-lam} approximately reproduces the Maxwell stress in the laminar state, it is not available for the turbulent state.

On the other hand, the fully developed turbulent state is empirically given from the data without electron heating.
We find an empirical formula of $\alpha_{\rm M, turb}$ from Table 3,
\begin{equation}\label{eq:turbeq}
	\alpha_{\rm M, turb} \approx 0.036  \pf{   \beta_{0}  }{10^{4} }^{-0.56} \ ,
\end{equation}
which can well reproduce $\alpha_{\rm M}$ of the case without electron heating in calculations of this paper. 

\begin{figure*}[t]
\centering
    \includegraphics[width=0.49\hsize,clip]{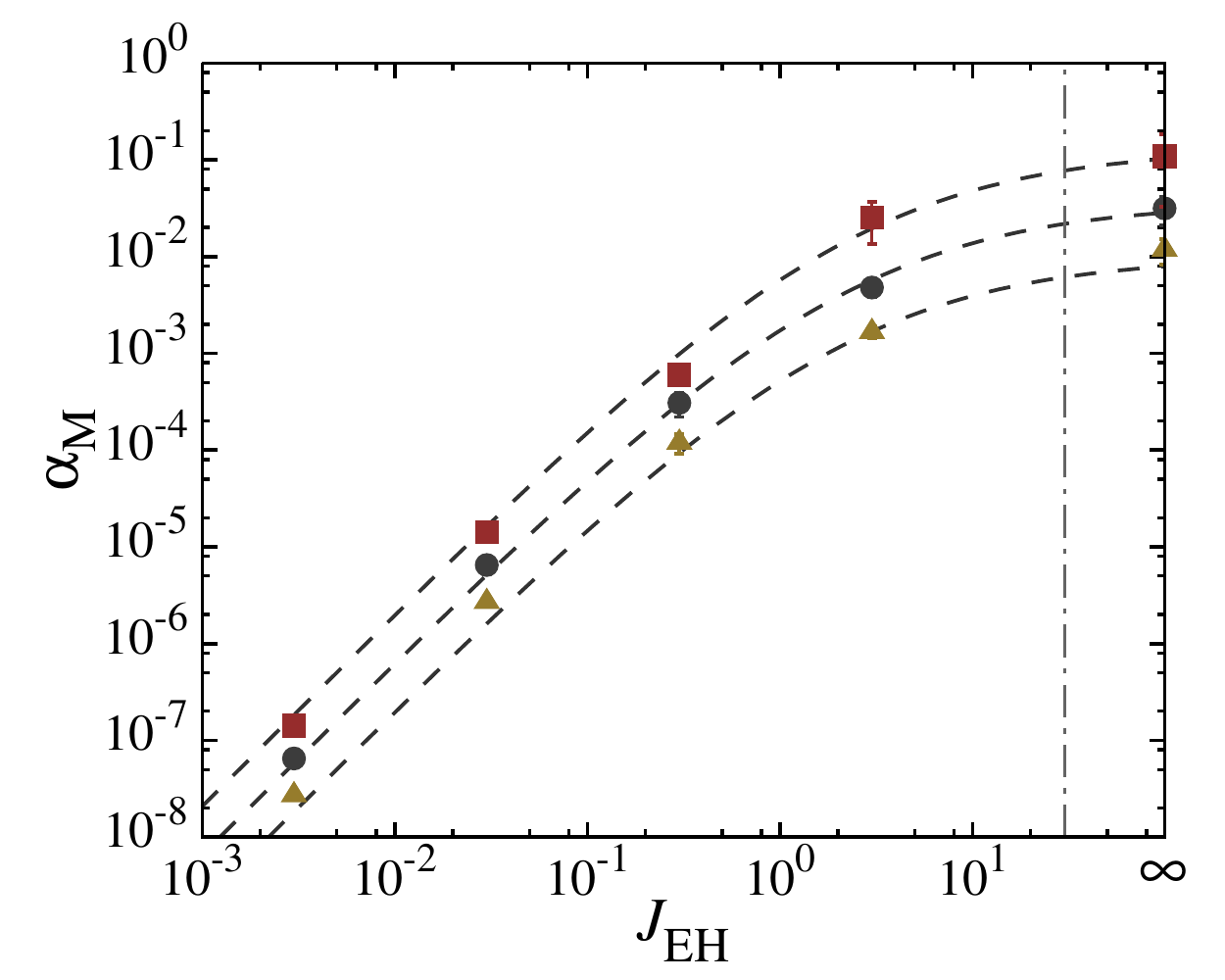} 
    \includegraphics[width=0.49\hsize,clip]{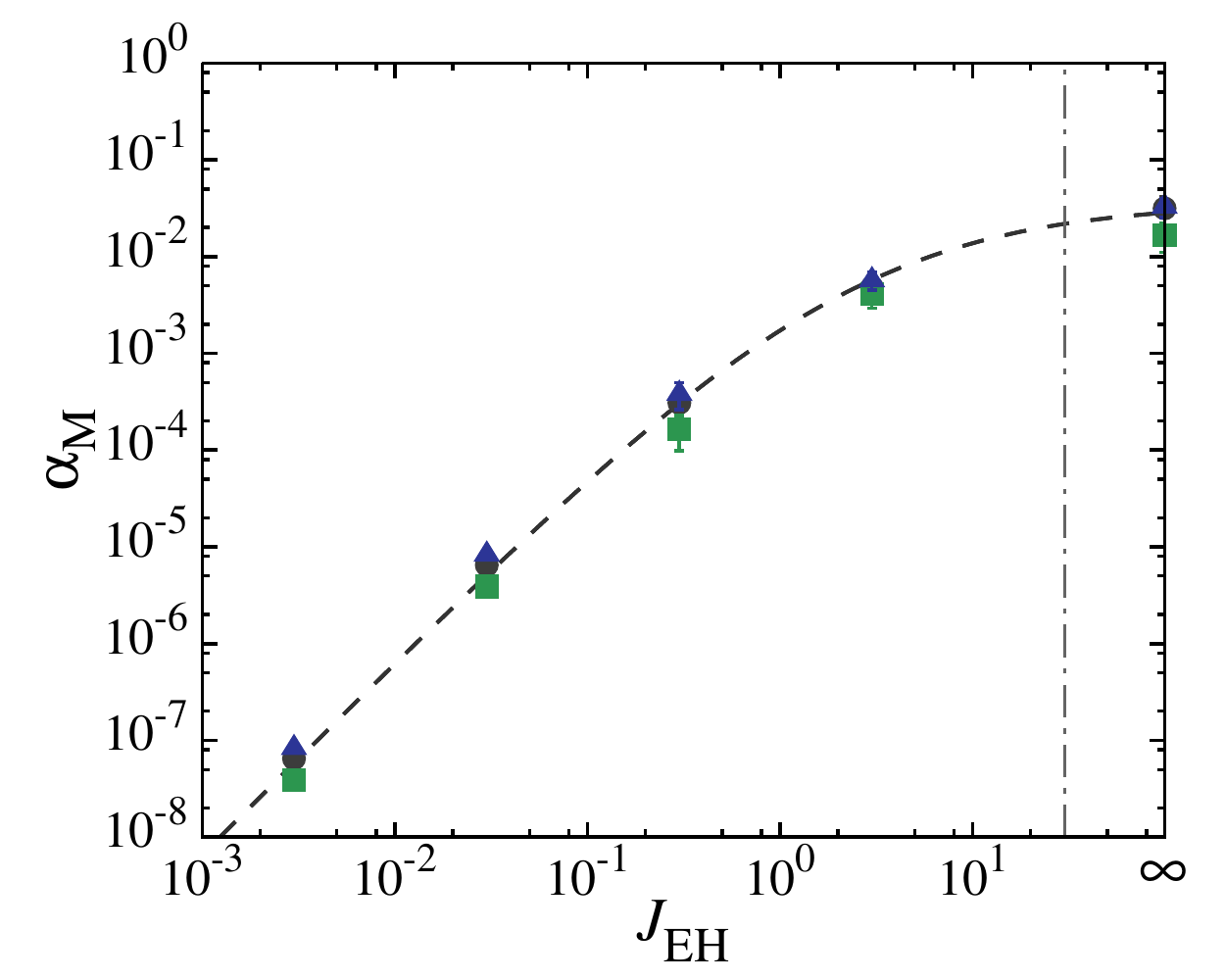} 
\caption{ The $\alpha_{\rm M}$ parameter as a function of  $J_{\rm EH}$ with varying initial plasma beta $\beta_{0}$ (left panel) and initial Elsasser number $\Lambda_{0}$ (right panel). In left panel, we show results for calculations with $\beta_{0}=10^{3}$ (red squares), with $\beta_{0}=10^{4}$ (black circles), and with $\beta_{0}=10^{5}$ (yellow triangles). In right panel, we show results for calculations with $\Lambda_{0}=30$ (blue triangles), with $\Lambda_{0}=10$ (black circles), and with $\Lambda_{0}=1$ (green squares). The dashed lines is fit by \eqref{eq:turb-lam}.}
 \label{fig:a-j-br}
\end{figure*}

To well reproduce simulation results, we make a function which approaches $\alpha_{\rm M, \,turb}$ and $\alpha_{\rm M, \,lam}$ with high $J_{\rm EH}$ limit and low $J_{\rm EH}$ limit, respectively, 
\begin{eqnarray}\label{eq:turb-lam}
	\alpha_{\rm M}  =  \pr{  \alpha_{\rm M, \,turb}^{-1/3}      +   \alpha_{\rm M, \, lam} ^{-1/3} }^{-3} \ .
\end{eqnarray}
To verify this equation, we compare them to the results with different $\beta_{0}$ ($\beta_{0} = 10^{3}, 10^{4},$ and $10^{5}$) and $\Lambda_{0}$ ($\Lambda_{0}= 30, 10,$ and $1$).
Figure \ref{fig:a-j-br} shows $\alpha_{\rm M}$--$J_{\rm EH}$ relation, with varying $\beta_{0}$ and $\Lambda_{0}$, respectively.
We see that \eqref{eq:turb-lam} well reproduce the resulting $\alpha_{\rm M}$.

We have to note that these results are based just on the simple analytic $J$--$E'$ relation.
In general, the saturated current density might not be equal to $J_{\rm EH}$.
In that case, the saturated current density would be required to be modified instead of $J_{\rm EH}$.
Moreover, the $J$--$E'$ relation including the electron heating can be multivalued function of $J$ \citepalias[see Figure 4 in][]{Mori2016Electron-Heatin}.
The electric fields may jump to the other blanch at $d J/ d E < 0$ because the electric field can vary with a much shorter timescale than the current density \citepalias[see more details in][]{Okuzumi2015The-Nonlinear-O}.
In that situation, the current density may not converge on a value at the final state.
This issue needs to be addressed in future calculations.

\section{Summary and Discussion}\label{sec:Summary}
We had investigated an effect of the electron heating on MRI, which has a potential to stabilize MRI  \citepalias{Okuzumi2015The-Nonlinear-O}. 
In this paper, we have performed the MHD simulation including the effect of damping a resistivity by the electron heating
to numerically show the possibility and efficiency of the electron heating.
We have clearly found that the electron heating suppresses the generation of the magnetic turbulence.
In particular, when the electron heating effectively operates, the ordered magnetic fields make the laminar flow.
The accretion stress caused by the magnetic fields is much less than the conventional turbulent stress of magnetic turbulence.
We also find a clear relation between the Maxwell stress and current density.
As the saturated current density is suppressed at lower and lower level by electron heating, the Maxwell stress becomes small.
Additionally, we have shown the analytical expression of the laminar flow, which allows us to predict the Maxwell stress in the presence of electron heating. 

The laminar flow formed by electron heating would have impacts on planetesimal formation.
As we see in Figure \ref{fig:sim-dif}, the vertical velocity dispersion drops when the electron heating completely suppresses the turbulence.
In the laminar flow, the turbulent diffusion in the vertical direction is no longer effective. 
Under the classical planetesimal formation theories, the dust sedimentation forms a dusty layer on midplane that might be gravitationally unstable \citep{Safronov1972Evolution-of-th,Goldreich1973The-Formation-o}.
The dust layer might cause the gravitational instability that forms planetesimals. 
This model has been focused in terms of avoiding the meter-size barrier.
However, vigorous disk turbulence easily stirs up the dust layer and diffuses it.
The dust layer with weak turbulence may also provide a possible place for secular gravitational instability that produces multiple ring-like structures and resulting planetesimals \citep{Takahashi2014Two-component-S,Takahashi2016An-Origin-of-Mu,Tominaga2017inprep}.
Therefore, weak disk turbulence may help the planetesimal formation.  
Such a dust sedimentation on midplane also help to cause the streaming instability which require high dust-to-gas mass ratio \citep{Youdin2005Streaming-Insta,Johansen2007Protoplanetary-,Bai2010Dynamics-of-Sol,Carrera2015How-to-form-pla}.
Therefore, efficient electron heating may help the formation of a dust layer and planetesimal formation. 
Moreover, such weak turbulent disk might explain observed disks suggested to be weak turbulence \citep[e.g.][]{Pinte2016Dust-and-Gas-in,Flaherty2017A-Three-Dimensi}.

In this paper, we neglect the stratified structure, non-Ohmic resistivities, and the negative slope in $J$--$E'$ relation predicted by the ionization calculation.
The stratified structure would affect the structure of magnetic field in the saturated state.
The non-Ohmic resistivities such as Hall effect and ambipolar diffusion would affect the final structure \citep[e.g.,][]{Bai2011Effect-of-Ambip,Kunz2013Magnetic-self-o,Lesur2014Thanatology-in-,Bethune2016Self-organisati,Bai2017Global-Simulati}, and therefore the importance of electron heating should be investigated with all resistivities. 
Moreover, the change of ionization balance by the electron heating would affect also the non-Ohmic resistivities. 
Although the simple analytic $J$--$E'$ relation could not address how much the current density would be saturated in reality, 
this work have shown that current density is suppressed by electron heating and there is the relation between Maxwell stress and current density.
We will address the saturated current density with the more detailed $J$--$E'$ relation in future work.

\acknowledgments 
The authors thank the anonymous referee for comments that improved the paper.
This work was supported by JSPS KAKENHI Grant Number JP15H02065, JP16K17661, JP16H04081, JP17J10129.
Numerical computations were carried out on Cray XC30 at Center for Computational Astrophysics, National Astronomical Observatory of Japan.

\end{document}